\documentclass[review]{elsarticle}

\usepackage{lineno,hyperref,mathtools}
\usepackage[letterpaper,top=2cm,bottom=2cm,left=3cm,right=3cm,marginparwidth=1.75cm]{geometry}

\journal{Journal of \LaTeX\ Templates}

\usepackage{amssymb}
\usepackage{bm}
\usepackage{amssymb}
\usepackage{amsthm}
\usepackage{multirow,booktabs}
\usepackage{cases}
\usepackage{threeparttable}

\newcommand{\F}{\mathbb{F}}

\newcommand{\C}{\mathcal{C}}
\newcommand{\lcm}{\mathrm{lcm}}

\newcommand{\MinRep}{\mathrm{MinRep}}

\newtheorem{theorem}{Theorem}

\newtheorem{remark}{Remark}

\newtheorem{lemma}{Lemma}
\newtheorem{corollary}{Corollary}
\newtheorem{example}{Example}
\begin{document}
	
	\begin{frontmatter}
		
		\title{A class of LCD BCH codes of length $n=\frac{q^{m}+1}{\lambda}$}
		\tnotetext[mytitlenote]{}
		
		\author[mymainaddress]{Yanhui Zhang}
		\ead{zhangyanhui0327@163.com}

	\address[mymainaddress]{School of Mathematics, Hefei University of Technology, Hefei 230009, China}
		\begin{abstract}
			LCD BCH codes are an important class of cyclic codes which have efficient encoding and decoding algorithms, but their parameters are difficult to determine. The objective of this paper is to study the LCD BCH codes of length $n=\frac{q^{m}+1}{\lambda}$, where $\lambda\mid (q+1)$ is an integer. Several types of LCD BCH codes with good parameters are presented, and many optimal linear codes are settled. Moreover, we present the first few largest coset leaders modulo $n=q^{m}+1, \frac{q^{m}+1}{2},\frac{3^{m}+1}{4}$, and partially solve two conjectures about BCH codes.
		\end{abstract}
		
		
	\end{frontmatter}
	\section{Introduction}
	Throughout this paper, let $p$ be a prime and $\F_{q}$ be a finite field, where $q>1$ is a $p$-power. An $[n,k,d]$ linear code over $\F_q$ is a linear subspace of $\F_q^{n}$ with dimension $k$ and minimum Hamming distance $d$. If $(c_{0}, c_{1},\ldots, c_{n-1}) \in \C$ implies $(c_{n-1}, c_{0}, c_{1},\ldots, c_{n-2})\in \C$, then $\C$ is called an $[n,k]$ cyclic code. Identify any vector $(c_{0}, c_{1},\ldots, c_{n-1})\in \F_q^{n}$ with
	\begin{center}
		$c_{0}+c_{1}x+c_{2}x^{2}+\cdots+c_{n-1}x^{n-1}\in \F_q[x]/ \langle x^{n}-1\rangle$,
	\end{center}
	then a cyclic code $\C$ over $\F_q$ is equivalent to an ideal of $\F_q[x]/\langle x^{n}-1\rangle$. 
	Note that $\F_q[x]/\langle x^{n}-1\rangle$ is a principal ideal ring, which means that every cyclic code $\C$ can be expressed as $\C$=$\langle g(x) \rangle$. If 
	$g(x)$ is the smallest degree monic polynomial such that $g(x)|(x^{n}-1)$ and $\C$=$\langle g(x) \rangle$, then $g(x)$ is called the generator polynomial, and $h(x) = (x^{n}-1)/g(x)$ is called check polynomial of $\C$. For a code $\C$, its (Euclidean) dual code is denoted as $\C^{\perp}:=\lbrace \mathbf{b}\in \F_{q}^{n}:\ \mathbf{b}\cdot\mathbf{c}^{T}=0 \ \text{for all}\ \mathbf{c}\in \C \rbrace$, where $^T$ denotes the transposition and $\cdot$ denotes the standard inner product. If $\C\cap\C^{\perp}=\{\mathbf{0}\}$, then code $\C$ is called an LCD code (linear code with complementary dual).
	
	Let $\F_{q^m}^*=\langle\alpha\rangle$ and $\beta= \alpha^{\frac{q^{m}-1}{n}}$, where $m=ord_n(q)$, then $\beta$ is a primitive $n$-th root of unity.
	Let $m_{i}(x)$ denote the minimal polynomial of $\beta ^{i}$ over $\F_q$, where $0\leq i\leq n-1$. For positive integers $b$ and $\delta$, define	
	\begin{center}
		$g_{(n,q,\delta,b)}(x):=\lcm \left(m_{b}(x),m_{b+1}(x),\ldots,m_{b+\delta -2}(x)\right)$,
	\end{center}
	where $2\leq \delta \leq n$, $\lcm$ denotes the least common multiple of these $m_i(x)$, $b\leq i\leq b+\delta-2$. Let $\C_{(n,q,\delta,b)}$ be the code over $\F_{q}$ with generator polynomial $g_{(n,q,\delta,b)}(x)$, then $\C_{(n,q,\delta,b)}$ is called a BCH code of length $n$ with designed distance $\delta$, and $\C_{(n,q,\delta,b)}$ is called a narrow-sense BCH code for $b=1$. When $n=q^{m}+1$, then $\C_{(n,q,\delta,b)}$ is called an antiprimitive BCH code in \cite{RefJ5}. Denote $d(\C_{(n,q,\delta,b)})$ as the minimum distance of $\C_{(n,q,\delta,b)}$ and $dim(\C_{(n,q,\delta,b)})$ as the dimension of $\C_{(n,q,\delta,b)}$, then $d(\C_{(n,q,\delta,b)})\geq\delta$ by employing the BCH bound.
	
	BCH codes were first proposed by Hocquenghem in \cite{RefJ8}, and then LCD codes were defined as reversible codes in \cite{RefJ18}. Although many scholars have obtained a lot of conclusions on BCH codes in \cite{RefJ1}-\cite{RefJ7}, \cite{RefJ12}-\cite{RefJ17} and \cite{RefJ20}-\cite{RefJ27}, but it is still hard to determine the parmeters of LCD BCH codes. Recently, the parameters of some LCD BCH codes of lengths $n$ were extensively studied in \cite{RefJ9},  \cite{RefJ19}, \cite{RefJ11}-\cite{RefJ16} and \cite{RefJ22}-\cite{RefJ25}, where $n=q^{m}-1,q^m+1,\frac{q^m-1}{q-1},\frac{q^m+1}{q+1}$, and  many researchers have developed great interest in the study of antiprimitive LCD BCH codes.
	
	In [\cite{RefJ12},\cite{RefJ13}, \cite{RefJ15},\cite{RefJ16},\cite{RefJ22},\cite{RefJ23},\cite{RefJ27}], the dimension of antiprimitive BCH code $\C_{(n,q,\delta,1)}$ was detemined for designed distance $2\leq \delta\leq q^ {\lceil\frac{m+2}{2}\rceil}+1$, and the first few largest coset leaders were given when $m=2t+1\geq 5$. However, it is hard to determine the first few largest coset leaders when $m=2t\geq2$, and some conjectures on the first few largest coset leaders were presented in \cite{RefJ16}. In \cite{RefJ27}, the authors gave a sufficient and necessary condition for
	$a$ being a coset leader, where $1\leq a\leq q^{m}$. Based on the above necessary and sufficient condition, the first few largest odd coset leaders were obtained in \cite{RefJ21}. In \cite{RefJ22}-\cite{RefJ24}, the dimension of LCD BCH code $\C_{(n,q,\delta,1)}$ with  $n=\frac{q^{m}+1}{2}$ and  $\frac{q^{m}+1}{q+1}$ has been studied.
	
	We shall work on the coset leaders modulo $n=\frac{q^{m}+1}{\lambda}$, and get the corresponding parameters of LCD BCH codes. The structure of this article is as follows: In section 2, we give some basic concepts and known results. In section 3, we provide the first few largest coset leaders when $\lambda=1$, and partially solve the conjectures in \cite{RefJ16}. In section 4, the dimensions of $\C_{(n,q,\delta,1)}$ and $\C_{(n,q,\delta+1,0)}$ are presented by employing $q$-cyclotomic cosets for $2\leq\lambda<q+1$, where $1\leq \delta-1\leq  \frac{2q^{t+1}+2q-1}{\lambda}$ if $2\nmid m$ and $\frac{q^t+1}{2}\leq \delta-1\leq \frac{q^{t+1}-1}{2}$ if $2\mid m$. In addition, we provide the dimensions of $\C_{(n,q,\delta,1)}$ and $\C_{(n,q,\delta+1,0)}$ for $\lambda=q+1$, where  $q^{t}+1\leq \delta-1\leq \frac{2q^{t+1}+2q-1}{q+1}$, and obtain the first two largest coset leaders modulo $n$ for $q=3$. In section 5, the conclusions of the paper are given.
	\section{Preliminaries}
	In this section, we give some basic concepts and known results that will be used later in this paper.
	\subsection{Basic Notations}
	For any $a$ with $0\leq a\leq q^m-1$, its $q$-adic expansion is  $a=\sum_{j=0}^{m-1}a_{j}q^{j}=(a_{m-1},a_{m-2},\ldots,a_{1},a_{0})$, where $0\leq a_{j}\leq q-1$ for $0\leq j\leq m-1$. Denote $\underbrace{A,A,\ldots,A}_{i\ times}$ as $\underbrace{A,\ldots}_{i}$ and $[u,v]:=\{u,u+1,\ldots,v\}$, where $u,v$ are integers with $u\leq v$.
	Let $T=\lbrace 0\leq i\leq n-1:g(\beta^{i})=0\rbrace$, where $\C_{(n,q,\delta,b)}$=$\langle g(x) \rangle$ and $\beta$ is a primitive $n$-th root of unity, then $T$ is referred as the defining set of $\C_{(n,q,\delta,b)}$. Denote $T^{-1}=\{n-t:i\in T\}$, then the  defining set of the dual code of $\C^{\perp}$ is $Z_{n}\textbackslash T^{-1}$ with respect to $\beta$.
	
	For any $s$ with $0\leq s\leq n-1$, the set
	\[\{sq^i\pmod{n}: \ 0\leq i\leq l-1\}\]is referred as the $q$-cyclotomic coset modulo $n$ of representative $s$ and is denoted by $C_{s}$, where $l=ord_{n}(q)$  is the multiplicative order of $q$ modulo $n$. The number of elements in $C_s$ is denoted by $|C_{s}|$. In addition, we denote $sq^i\pmod{n}$ by $[sq^{i}]_{n}$ for integer $0\leq i\leq l-1$. We assume that $CL(s):=\min\{i:\ i\in C_{s}\}$ and $\MinRep_{n}:=\{CL(s): \ 0\leq s\leq n-1\}$, then $CL(s)$ is denoted as the coset leader of $C_{s}$, and $\delta_{i,n}$ is denoted as the $i$-th largest coset leader modulo $n$.	
	
	\subsection{Known Results on Coset Leaders and the Dimension of BCH Code}
	For the BCH code of length $n=\frac{q^m+1}{\lambda}$, it is difficult to determine the coset leaders modulo $n$ and the dimension of $\C_{(n,q,\delta,b)}$. We list some known conclusions as follows.
	\begin{lemma}(\cite{RefJ27})\label{le1}
		Let $n=q^{m}+1$, $i$ and $h$ be integers such that $1\leq i< m$, 	$-\frac{l(q^{m-i}-1)}{q^{i}+1}<h<\frac{l(q^{m-i}+1)}{q^{i}-1}$. Then
		\begin{itemize}
			\item[(1)]If $2\mid q$ and $m\geq 2$, let 
			$1\leq l\leq \big\lfloor \frac{(q^{i}-1)q}{2(q+1)}\big\rfloor$ be an integer.
			Then $a$ is a coset leader modulo $n$ with $0\leq a\leq q^{m}$ if and only if $0\leq a\leq \lfloor\frac{q^{m+1}+q}{2(q+1)}\rfloor$ and $a\neq lq^{m-i}+h$.
			\item[(2)]If $2\nmid q$, let 
			$1\leq l\leq \frac{q^{i}-1}{2}$ be an integer.
			Then $a$ is a coset leader modulo $n$ with $0\leq a\leq q^{m}$ if and only if $a\leq \frac{q^{m}+1}{2}$ and $a\neq lq^{m-i}+h$.
		\end{itemize}
	\end{lemma}
	\begin{lemma}(\cite{RefJ23})\label{le2}
		Let $n=\frac{q^{m}+1}{\lambda}$ and $1\leq a\leq n-1$, where $2\nmid q$ and $\lambda\mid (q+1)$. Then $a$ is a coset leader of $C_{a}$ modulo $n$ if and only if $\lambda a$ is a coset leader of $C_{\lambda a}$ modulo $q^{m}+1$. Moreover, $|C_{a}|=|C_{\lambda a}|$.
	\end{lemma}
	\begin{lemma}(\cite{RefJ16})\label{le3}
		Let $n=q^{m}+1$ and $m\geq 4$ be an integer. Then  
		\begin{itemize}
			\item[(a)] If $2\nmid q$ and $2\nmid m$, then $$\delta_{1,n}=\frac{q^{m}+1}{2},\delta_{2,n}=\frac{(q-1)(q^{m}+1)}{2(q+1)},\delta_{3,n}=\frac{(q-1)(q^{m}-2q^{m-2}-1)}{2(q+1)},\delta_{4,n}=\delta_{3,n}-(q-1)^{2}.$$ Moreover, $|C_{\delta_{1,n}}|=1,|C_{\delta_{2,n}}| =2,| C_{\delta_{3,n}}| =|C_{\delta_{4,n}}| =2m.$
			\item[(b)]If $2\nmid q$ and $m=2^{r}t+2^{r-1}$, where $r\geq 2$ and $t\geq 1$ are integers. Let $\Phi (x,q)=\frac{q-1}{2}(q^{2^{x}}-1)(q^{2^{x-1}}-1)\cdots(q^{2^{0}}-1)$, where $x\geq 0$. Then
			$$\delta_{1,n}=\frac{q^{m}+1}{2},\quad \delta_{2,n}=\frac{n}{q^{2^{r-1}}+1}\cdot\Phi(r-2,q).$$ 
			Moreover, $|C_{\delta_{1,n}}|=1\ and \ |C_{\delta_{2,n}}|=2^{r}.$
			\item[(c)] If $2\mid q$ and $m=2^{r}t+2^{r-1}$, where $r\geq 2$ and $t\geq 1$ are integers. Let $\Phi^{'} (x,q)=\frac{q}{2}(q^{2^{x}}-1)(q^{2^{x-1}}-1)\cdots(q^{2^{0}}-1)$, where $x\geq 0$. Then
			$$\delta_{1,n}=\frac{n}{q^{2^{r-1}}+1}\cdot\Phi^{'}(r-2,q).$$ 
		\end{itemize}
	\end{lemma}
	\begin{lemma}(\cite{RefJ12}\cite{RefJ16})\label{le4}
		Let $n=q^{m}+1$. Then 
		\begin{itemize}	
			\item[(a)] If $m\geq3$ is an odd integer and $1\leq a\leq q^{\frac{m+1}{2}}$ with $a\not\equiv0\pmod{q}$, then $a$ is a coset leader with $|C_{a}|=2m$ except that $a=q^{\frac{m+1}{2}}-c $ for $1\leq c\leq q-1$.
			\item[(b)] If $2\nmid q$ and $m\geq4$ is integer. Then
			\begin{itemize}
				\item[(b.1)] If $m=2t+1$ and $1\leq a< 2q^{t+1}-2q+1$ with $a\not\equiv0\pmod{q}$, then $a$ is a coset leader and $|C_{a}| =2m$ except that
				\begin{center}
					$a=q^{t+1}\pm \gamma$ or	$a=q^{t+1}+\alpha q^{t}\pm 1$,
				\end{center}
				where $\gamma, \alpha\in [1,q-1]$.
				\item[(b.2)] If $m=2t+1$ and $2q^{t+1}-2q+1\leq a\leq 2q^{t+1}+2q-1$, then $a$ is not a coset leader.
				\item[(b.3)]If $m=2t$ and $1\leq a\leq 2q^{t}+2$ with $a\not\equiv0\pmod{q}$, then $a$ is a coset leader and $|C_{a}| =2m$ except that
				\begin{center}
					$a=q^{t}+1$ or $a=2q^{t}\pm 1$ or $a=2q^{t}+2$.
				\end{center}
			\end{itemize}
		\end{itemize}
	\end{lemma}
	\begin{lemma}\label{le5}
		Let $n=\frac{q^{m}+1}{\lambda}$, where $m\geq 2$ and $\lambda \mid (q+1)$ with $q+1>\lambda$, then we have $ord_{n}(q)=2m$. Moreover, let $C_{s}$ be a cyclotomic coset modulo $n$ containing $s$, then $C_{s}=\{y_{s,k},n-y_{s,k}: 0\leq k\leq m-1\}$, where $y_{s,k}=sq^k\pmod{n}$.
		\begin{proof}
			It is clear that $n\mid (q^{2m}-1)$, we denote $l=ord_{n}(q)$, then $l\mid 2m$. Since $\frac{q+1}{\lambda}\geq 2$, then $n=\frac{q^{m}+1}{\lambda}=\frac{q+1}{\lambda}\frac{q^{m}+1}{q+1}\geq 2\frac{q^{m}+1}{q+1}>q^{m-1}+1$, then $l\geq m$ by $n\mid (q^{l}-1)$. If $l=m$, then $n\mid (q^{m}-1)$. Note that $gcd(q^{m}+1,q^{m}-1)=1$ or $2$, this implies $n\mid 1$ or $n\mid 2$, which is impossible. Hence, $ord_{n}(q)=2m$.
			
			For any $s\in Z_{n}$, then $C_{s}=\lbrace y_{s,0},y_{s,1},\ldots,y_{s,k},\ldots,y_{s,2m-1}\rbrace$ by $ord_{n}(q)=2m$. If $m\leq k \leq 2m-1$, then $y_{s,k}\equiv sq^{k}=s(q^{m}+1-1)q^{k-m}\equiv -sq^{k-m}\equiv n-sq^{k-m}$. Therefore, $C_{s}$ can be denoted as $C_{s}=\{y_{s,k},n-y_{s,k} : 0\leq k\leq m-1\}$. Thus we complete the proof.
		\end{proof}
	\end{lemma}
	Let $n=\frac{q^{m}+1}{\lambda}$, where $m\geq 2$ and $\lambda \mid (q+1)$ with $q+1>\lambda$, we have $T\cap Z_{n}\textbackslash T^{-1}=\emptyset$ for BCH code $\C_{(n,q,\delta,b)}$ by Lemma \ref{le5}. Therefore, we have $\C\cap \C^{\perp}=\{\mathbf{0}\}$, i.e., $\C_{(n,q,\delta,b)}$ is an LCD BCH code. Moreover, we have known that BCH code of length $n=\frac{q^m+1}{q+1}$ is an LCD BCH code in \cite{RefJ22}.
	\section{The LCD BCH Codes of Length $n=q^{m}+1$}
	In this section, we always assume that $n=q^{m}+1$ with $m=4t+2\geq 2$. This section will be divided into two subsections by the value of $q$. In the following, we will investigate the first few largest coset leaders $\delta_{i,n}$ and the parameters of $\C_{(n,q,\delta_{i,n},b)}$. 
	\subsection{When $q$ is an even prime power}
	\begin{lemma}\label{le6}
		Let $q$ be an even prime power and $m\equiv2 \pmod{4}$ with $m\geq 6$, then $	\delta_{1,n}=\frac{q(q-1)(q^{m}+1)}{2(q^{2}+1)}$ and $|C_{\delta_{1,n}}|=4$.
		\begin{proof}
			Note that $m=2^{2}t+2$, then $\delta_{1,n}=\frac{q(q-1)(q^{m}+1)}{2(q^{2}+1)}$ by Lemma \ref{le3}. It is clear that 
			$$[\delta_{1,n}q]_{n}=\frac{(q^{2}-q+2)(q^{m}+1)}{2(q^{2}+1)},\quad [\delta_{1,n}q^{2}]_{n}=\frac{(q^{2}+q+2)(q^{m}+1)}{2(q^{2}+1)},$$
			$$[\delta_{1,n}q^{3}]_{n}=\frac{(q^{2}+q)(q^{m}+1)}{2(q^{2}+1)},\quad [\delta_{1,n}q^{4}]_{n}=\frac{q(q-1)(q^{m}+1)}{2(q^{2}+1)}=\delta_{1,n},$$
			then $|C_{\delta_{1,n}}|=4$. Thus we complete the proof.
		\end{proof}
	\end{lemma}
	\begin{lemma}\label{le7}
		Let $q$ be an even prime power and $m\equiv2 \pmod{4}$ with $m\geq 6$, then 
		\begin{center}
			$\delta_{2}=\frac{q(q-1)(q^{m}-2q^{m-4}-1)}{2(q^{2}+1)}$ and 
			$\delta_{3}=$
			$\begin{cases}
				\delta_{2}-(q-1)q^{2}, &\text{if}\ m=6;\\
				\delta_{2}-q(q-1)(q^{2}-1), &\text{if}\ m>6.	
			\end{cases}
			$
		\end{center} 
		are the second and third largest coset leaders modulo $n$. Moreover, $|C_{\delta_{2}}|=|C_{\delta_{3}}|=2m$.
		\begin{proof}
			From Lemma \ref{le1}, we have $a\in \MinRep_{n}$ with $0\leq a\leq \lfloor\frac{q^{m+1}+q}{2(q+1)}\rfloor$ which implies that there are no integers $1\leq i\leq m-1$, $1\leq l\leq \big\lfloor \frac{(q^{i}-1)q}{2(q+1)}\big\rfloor$ and $-\frac{l(q^{m-i}-1)}{q^{i}+1}<h<\frac{l(q^{m-i}+1)}{q^{i}-1}$ such that 
			\begin{equation}\label{eq1}
				a=lq^{m-i}+h.
			\end{equation}
			If (\ref{eq1}) holds, then
			$$\frac{l(q^{m}+1)}{q^{i}+1}<a<\frac{l(q^{m}+1)}{q^{i}-1},$$
			which is equivalent to 
			\begin{equation}\label{eq2}
				\frac{a(q^{i}-1)}{q^{m}+1}<l<\frac{a(q^{i}+1)}{q^{m}+1}.
			\end{equation}
			For any $1\leq i\leq m-1$, suppose that $$aq^{i}=u_{i}(q^{m}+1)+[aq^{i}]_{n}.$$
			Hence, (\ref{eq2}) becomes
			\begin{equation}\label{eq3}
				u_{i}+\frac{[aq^{i}]_{n}-a}{q^{m}+1}<l<u_{i}+\frac{[aq^{i}]_{n}+a}{q^{m}+1}.
			\end{equation}
			Therefore, if there is no $l$ such that (\ref{eq3}) holds for any $1\leq i\leq m-1$, then this implies that $a\in \MinRep_{n}$.
			In order to prove our results, we divide it into two steps.
			
			$\bf{Step 1}$. We claim that  $\delta_{2}$ and $\delta_{3}$ are coset leaders. 
			\\$\bf{(1)}$ When $m>6$, we have $$\delta_{2}=\frac{q(q-1)(q^{m}-2q^{m-4}-1)}{2(q^{2}+1)}=\frac{q(q-1)}{2}\big(\frac{q^{m}+1}{q^{2}+1}-2\frac{q^{m-4}+1}{q^{2}+1}\big)<\big\lfloor\frac{q^{m+1}+q}{2(q+1)}\big\rfloor.$$
			Note that $\frac{q^{m}+1}{q^{2}+1}=1+\sum_{i=0}^{\frac{m-6}{4}}[(q-1)q+(q-1)]q^{4i+2}$, then
			\begin{center}
				$\begin{aligned}
					\delta_{2}&=\frac{q(q-1)}{2}\big([q(q-1)+(q-1)]q^{m-4}-1-\sum_{i=0}^{\frac{m-6}{4}-1}[(q-1)q+(q-1)]q^{4i+2}\big)\\
					&=(\frac{q}{2}-1,\frac{q}{2}-1,\frac{q}{2},\frac{q}{2}-1,\underbrace{\underbrace{\frac{q}{2},\frac{q}{2},\frac{q}{2}-1,\frac{q}{2}-1},\ldots}_{\frac{m-6}{4}},\frac{q}{2},\frac{q}{2}).
				\end{aligned}$
			\end{center}
			$\bf{(1.1)}$ For $1\leq i\leq 3$, we have
			$$\left\{\begin{aligned}
				[\delta_{2}q]_{n}=&(\frac{q}{2}-1,\frac{q}{2},\frac{q}{2}-1,\underbrace{\underbrace{\frac{q}{2},\frac{q}{2},\frac{q}{2}-1,\frac{q}{2}-1},\ldots}_{\frac{m-6}{4}},\frac{q}{2},\frac{q}{2}-1,\frac{q}{2}+1)>\delta_{2},\\
				[\delta_{2}q]_{n}+\delta_{2}=&(q-1,0,0,\underbrace{\underbrace{0,0,q-1,q-1},\ldots}_{\frac{m-6}{4}},0,0,1)<q^{m}+1.
			\end{aligned}
			\right.$$
			\begin{center}
				$\left\{\begin{aligned}
					[\delta_{2}q^{2}]_{n}=&(\frac{q}{2},\frac{q}{2}-1,\underbrace{\underbrace{\frac{q}{2},\frac{q}{2},\frac{q}{2}-1,\frac{q}{2}-1},\ldots}_{\frac{m-6}{4}},\frac{q}{2},\frac{q}{2}-1,\frac{q}{2},\frac{q}{2}+1)>\delta_{2},\\
					[\delta_{2}q^{2}]_{n}+\delta_{2}=&(q-1,q-1,0,\underbrace{q-1,\ldots}_{m-6},q-1,1,1)<q^{m}+1.
				\end{aligned}
				\right.$
			\end{center}
			\begin{center}
				$\left\{\begin{aligned}
					[\delta_{2}q^{3}]_{n}=&(\frac{q}{2}-1,\underbrace{\underbrace{\frac{q}{2},\frac{q}{2},\frac{q}{2}-1,\frac{q}{2}-1},\ldots}_{\frac{m-6}{4}},\frac{q}{2},\frac{q}{2}-1,\frac{q}{2},\frac{q}{2},\frac{q}{2})>\delta_{2},\\
					[\delta_{2}q^{3}]_{n}+\delta_{2}=&(q-1,0,0,q-1,\underbrace{\underbrace{0,0,q-1,q-1},\ldots}_{\frac{m-10}{4}},0,0,q-1,0,1,0)<q^{m}+1.
				\end{aligned}
				\right.$
			\end{center}
			$\bf{(1.2)}$ For $4\leq i\leq m-3$, we assume $1\leq k\leq \frac{m-6}{4}$.
			\begin{itemize}
				\item[$\bf{(i)}$] For $[\delta_{2}q^{4k}]_{n}$, we have
				\begin{center}
					$[\delta_{2}q^{4k}]_{n}=$
					$\left\{\begin{aligned}
						&(\underbrace{\underbrace{\frac{q}{2},\frac{q}{2},\frac{q}{2}-1,\frac{q}{2}-1},\ldots}_{\frac{m-6}{4}},\frac{q}{2},\frac{q}{2}-1,\frac{q}{2},\frac{q}{2},\frac{q}{2}-1,\frac{q}{2}+1), \quad or\\
						&(\frac{q}{2},\frac{q}{2},\frac{q}{2}-1,\frac{q}{2}-1,\ldots,\frac{q}{2},\frac{q}{2}-1,\frac{q}{2},\frac{q}{2},\frac{q}{2}-1,\frac{q}{2},\ldots,\frac{q}{2}-1,\frac{q}{2}-1,\frac{q}{2},\frac{q}{2}+1).
					\end{aligned}
					\right.$
				\end{center}
				Clearly, $\delta_{2}<[\delta_{2}q^{4k}]_{n}$. Note that $\delta_{2}<(\frac{q}{2}-1,\frac{q}{2}-1,\frac{q}{2},\frac{q}{2},0,\underbrace{0,\ldots}_{m-6},0)$ and $[\delta_{2}q^{4k}]_{n}<(\frac{q}{2},\frac{q}{2},\frac{q}{2}-1,\frac{q}{2}-1,\frac{q}{2},\underbrace{\frac{q}{2},\ldots}_{m-6},\frac{q}{2})$, then  $$[\delta_{2}q^{4k}]_{n}+\delta_{2}<(q-1,q-1,q-1,q-1,\frac{q}{2},\underbrace{\frac{q}{2},\ldots}_{m-6},\frac{q}{2})<q^{m}+1.$$
				\item[$\bf{(ii)}$] For $[\delta_{2}q^{4k+1}]_{n}$, we have
				\begin{center}
					$[\delta_{2}q^{4k+1}]_{n}=$
					$\left\{\begin{aligned}
						&(\frac{q}{2},\frac{q}{2}-1,\frac{q}{2}-1,\underbrace{\underbrace{\frac{q}{2},\frac{q}{2},\frac{q}{2}-1,\frac{q}{2}-1},\ldots}_{\frac{m-10}{4}},\frac{q}{2},\frac{q}{2}-1,\frac{q}{2},\frac{q}{2},\frac{q}{2}-1,\frac{q}{2},\frac{q}{2}),\quad or\\
						&(\frac{q}{2},\frac{q}{2}-1,\frac{q}{2}-1,\ldots,\frac{q}{2},\frac{q}{2}-1,\frac{q}{2},\frac{q}{2},\frac{q}{2}-1,\frac{q}{2},\ldots,\frac{q}{2}-1,\frac{q}{2}-1,\frac{q}{2},\frac{q}{2},\frac{q}{2}).
					\end{aligned}
					\right.$
				\end{center}
				Clearly, $\delta_{2}<[\delta_{2}q^{4k+1}]_{n}$. Note that $[\delta_{2}q^{4k+1}]_{n}<[\delta_{2}q^{4k}]_{n}$, then  $[\delta_{2}q^{4k+1}]_{n}+\delta_{2}<q^{m}+1.$
				\item[$\bf{(iii)}$]For $[\delta_{2}q^{4k+2}]_{n}$ and $[\delta_{2}q^{4k+3}]_{n}$, we have $\delta_{2}<[\delta_{2}q^{4k+2}]_{n},[\delta_{2}q^{4k+3}]_{n}$ and $[\delta_{2}q^{4k+2}]_{n}+\delta_{2},[\delta_{2}q^{4k+3}]_{n}+\delta_{2}<q^{m}+1$ by the same way of $\bf{(ii)}$.
			\end{itemize}
			$\bf{(1.3)}$ For $i=m-2,m-1$, we have
			$$\begin{cases}
				[\delta_{2}q^{m-2}]_{n}=\big(\frac{q}{2},\frac{q}{2}-1,\frac{q}{2},\frac{q}{2},\frac{q}{2}-1,\frac{q}{2},\ldots,\frac{q}{2},\frac{q}{2}+1\big)>\delta_{2},\\
				[\delta_{2}q^{m-1}]_{n}=(\frac{q}{2}-1,\frac{q}{2},\frac{q}{2},\frac{q}{2}-1,\frac{q}{2},\frac{q}{2}-1,\ldots,\frac{q}{2},\frac{q}{2})>\delta_{2}.
			\end{cases}$$
			Note that $[\delta_{2}q^{m-2}]_{n},[\delta_{2}q^{m-1}]_{n}<[\delta_{2}q^{4k}]_{n}$, then  $[\delta_{2}q^{m-2}]_{n}+\delta_{2},[\delta_{2}q^{m-1}]_{n}+\delta_{2}<q^{m}+1.$
			
			Concluding all discussions above, we have $0<\frac{[\delta_{2}q^{i}]_{n}-\delta_{2}}{q^{m}+1}<\frac{[\delta_{2}q^{i}]_{n}+\delta_{2}}{q^{m}+1}<1$ for any $1\leq i\leq m-1$. Hence, there is no $l$ such that (\ref{eq3}) holds, this implies that $\delta_{2}\in \MinRep_{n}$.
			On one hand, $\delta_{2}<[\delta_{2}q^{i}]_{n}$ for $1\leq i\leq m-1$, and 
			$$[\delta_{2}q^{m}]_{n}=\big(\frac{q}{2},\frac{q}{2},\frac{q}{2}-1,\frac{q}{2},\frac{q}{2}-1,\frac{q}{2}-1,\ldots,\frac{q}{2}-1,\frac{q}{2}+1\big)>\delta_{2},$$
			then $|C_{\delta_{2}}| >m$. On the other hand, $|C_{\delta_{2}}|$ is a divisor of $2m$, then $|C_{\delta_{2}}| =2m$.
			\\Note that 
			$$\begin{aligned}
				\delta_{3}&=\delta_{2}-q(q-1)(q^{2}-1)\\
				&=(\frac{q}{2}-1,\frac{q}{2}-1,\frac{q}{2},\frac{q}{2}-1,\underbrace{\underbrace{\frac{q}{2},\frac{q}{2},\frac{q}{2}-1,\frac{q}{2}-1},\ldots}_{\frac{m-10}{4}},\frac{q}{2},\frac{q}{2}-1,\frac{q}{2},\frac{q}{2},\frac{q}{2}-1,\frac{q}{2}).
			\end{aligned}$$
			Then the proof of $\delta_{3}\in \MinRep_{n}$ and $|C_{\delta_{3}}| =2m$ are similar to $\delta_{2}$, and details are omitted here.
			\\$\bf{(2)}$ When $m=6$, we have 
			\begin{center}
				$\begin{cases}
					\delta_{2}=(\frac{q}{2}-1,\frac{q}{2}-1,\frac{q}{2},\frac{q}{2}-1,\frac{q}{2},\frac{q}{2}),\\
					\delta_{3}=(\frac{q}{2}-1,\frac{q}{2}-1,\frac{q}{2}-1,\frac{q}{2},\frac{q}{2},\frac{q}{2}).
				\end{cases}$
			\end{center}
			We can calculate that $\delta_{2}\leq[\delta_{2}q^{i}]_{n}$, $n-[\delta_{2}q^{i}]_{n}$ and $\delta_{3}\leq [\delta_{3}q^{i}]_{n}$, $n-[\delta_{3}q^{i}]_{n}$ for any $0\leq i\leq 5$, then 
			$\delta_{2}$, $\delta_{3}\in \MinRep_{n}$ and $|C_{\delta_{2}}|=|C_{\delta_{3}}|=2m$.
			
			$\bf{Step 2}$. We claim that $\delta_{2,n}=\delta_{2}$ and $\delta_{3,n}=\delta_{3}$. 
			\\$\bf{(1)}$ When $m>6$, we assume that there is an integer $a$ such that $a\in \MinRep_{n}$ and $\delta_{2}<a<\delta_{1,n}$. From Lemma \ref{le6}, we have 
			$$\delta_{1,n}=\frac{q(q-1)(q^{m}+1)}{2(q^{2}+1)}=\frac{q(q-1)}{2}\frac{q^{m}+1}{q^{2}+1}=(\underbrace{\underbrace{\frac{q}{2}-1,\frac{q}{2}-1,\frac{q}{2},\frac{q}{2}},\ldots}_{\frac{m-2}{4}},\frac{q}{2}-1,\frac{q}{2}).$$
			Let $a=\sum_{i=0}^{m-1}a_{i}q^{i}$, where $0\leq a_{i}\leq q-1$. Note that $a\in \MinRep_{n}$ and $\delta_{2}<a<\delta_{1,n}$, then $a_{m-1}=\frac{q}{2}-1,a_{m-2}=\frac{q}{2}-1,a_{m-3}=\frac{q}{2},a_{m-4}=\frac{q}{2}-1$ or $\frac{q}{2},a_{0}\neq 0$.
			\\$\bf{(1.1)}$ Suppose there exists an integer $k$ such that $a_{k}\notin\{\frac{q}{2}-1,\frac{q}{2} \}$, where $0\leq k\leq m-5$.
			\begin{itemize}
				\item[$\bf{(i)}$] If $a_{k}\leq \frac{q}{2}-2$, then $[aq^{m-k-1}]_{n}<(\frac{q}{2}-2,q-1,q-1,\ldots,q-1)<\delta_{2}<a$, which is a contradiction to the assumption of
				$a\in \MinRep_{n}$.
				\item[$\bf{(ii)}$] If $a_{k}\geq \frac{q}{2}+1$, then
				\begin{center}
					$\begin{cases}
						[aq^{m-k-1}]_{n}>(\frac{q}{2},\frac{q}{2},\frac{q}{2},\frac{q}{2}-1,0,\ldots,0),     &\text{if}\ k=0;\\
						[aq^{m-k-1}]_{n}>(\frac{q}{2}+1,0,0,\ldots,0),  &\text{if}\ k\neq 0.
					\end{cases}$
				\end{center}  Note that $n-[aq^{m-k-1}]_{n}\in C_{a}$, but $$n-[aq^{m-k-1}]_{n}<(\frac{q}{2}-1,\frac{q}{2}-1,\frac{q}{2}-1,\frac{q}{2},0,0,\ldots,0)+2<\delta_{2,n}<a,$$ which gives a contradiction to $a\in \MinRep_{n}$.
			\end{itemize}
			$\bf{(1.2)}$ For $a_{i}\in \{\ \frac{q}{2}-1,\frac{q}{2}\}$, where $0\leq i\leq m-5$. Let the $q$-expansion $\delta_{1,n}=\sum_{i=0}^{m-1}b_{i}q^{i}$ and $\delta_{2}=\sum_{i=0}^{m-1}c_{i}q^{i}$, where $0\leq b_{i},c_{i}\leq q-1$.
			\begin{itemize}
				\item[$\bf{(i)}$] If $a_{m-4}=\frac{q}{2}$, since $a<\delta_{1,n}$ then there exists an integer $k$ such that $a_{k}=\frac{q}{2}-1$, $b_{k}=\frac{q}{2}$ and $a_{i}=b_{i}$ for any $k<i\leq m-1$. 
				\\When $k=0$, then $$[aq^{m-1}]_{n}=(\frac{q}{2}-2,\underbrace{\underbrace{\frac{q}{2},\frac{q}{2},\frac{q}{2}-1,\frac{q}{2}-1},\ldots}_{\frac{m-2}{4}},\frac{q}{2}+1)<\delta_{2}<a.$$
				\\When $k\neq 0$, then we have $a_{k+2}=a_{k+1}=a_{k}=\frac{q}{2}-1$ or
				$\begin{cases}
					a_{k+3}=a_{k+2}=a_{k}=\frac{q}{2}-1\\
					a_{k+1}=\frac{q}{2}
				\end{cases}$. It is easy to check that $[aq^{m-k-3}]_{n}<a$ or $[aq^{m-k-4}]_{n}<a$, this contradicts to $a\in \MinRep_{n}$.
				\item[$\bf{(ii)}$]If $a_{m-4}=\frac{q}{2}-1$, since $\delta_{2}<a$ then there exists an integer $k$ such that $a_{k}=\frac{q}{2}$, $c_{k}=\frac{q}{2}-1$ and $a_{i}=c_{i}$ for any $k<i\leq m-1$. Since $c_{1}=c_{0}=\frac{q}{2}$, then $2\leq k$, then we have $a_{k+2}=a_{k+1}=a_{k}=\frac{q}{2}$ or 
				$\begin{cases}
					a_{k+3}=a_{k+2}=a_{k}=\frac{q}{2}\\
					a_{k+1}=\frac{q}{2}-1
				\end{cases}$. 
				It is easy to check that $n-[aq^{m-k-3}]_{n}<a$ or $n-[aq^{m-k-4}]_{n}<a$, this contradicts to the assumption of $a\in \MinRep_{n}$.
			\end{itemize}
			$\bf{(2)}$ When $m=6$, we have $$\delta_{1,n}=(\frac{q}{2}-1,\frac{q}{2}-1,\frac{q}{2},\frac{q}{2},\frac{q}{2}-1,\frac{q}{2}).$$ Suppose $a\in \MinRep_{n}$ and $\delta_{2}<a<\delta_{1,n}$, let $a=\sum_{i=0}^{m-1}a_{i}q^{i}$, where $0\leq a_{i}\leq q-1$. We can get $a_{i}\in\{\frac{q}{2}-1,\frac{q}{2}\}$ by the similar way of $\bf{(1.2)}$, then $a=(\frac{q}{2}-1,\frac{q}{2}-1,\frac{q}{2},\frac{q}{2},\frac{q}{2}-1,\frac{q}{2}-1)$. But $[aq^{5}]_{n}<a$, which contradicts to the assumption of $a\in \MinRep_{n}$.
			
			Collecting all discussions above, we have $\delta_{2,n}=\delta_{2}$ and we can prove $\delta_{3,n}=\delta_{3}$ similarly. Thus we complete the proof.
		\end{proof}
	\end{lemma}
	\begin{lemma}\label{le8}
		Let $n=q^{2}+1$ and $q$ be an even prime power. Then $a\in \MinRep_{n}$ if and only if $a\in [l(q+1)+1,(l+1)(q-1)]$, where $0\leq l\leq \frac{q}{2}-1$. Moreover, $|C_{a}|=4$ for any $a\in \MinRep_{n}$.
		\begin{proof}
			If $q\mid a$ and $a\neq 1$, we have $a\in C_{1}$, i.e., $a\notin\MinRep_{n}$. 
			For $1\leq a\leq q-1$, it is clear that $a<aq$, $a<n-aq$ and $a<n-a$, i.e., $a\in \MinRep_{n}$. 
			
			For $q+1\leq a\leq n-1$, let $a=a_{1}q+a_{0}$, where $1\leq a_{0},a_{1}\leq q-1$. Note that $$C_{a}=\{a,a_{0}q-a_{1},(q-a_{1})q-(a_{0}-1),(q-a_{0})q+a_{1}+1\}.$$
			If $a\in\MinRep_{n}$, then 
			\begin{center}
				$\begin{cases}
					a_{1}q+a_{0}\leq a_{0}q-a_{1},\\
					a_{1}q+a_{0}\leq (q-a_{1})q-(a_{0}-1),\\
					a_{1}q+a_{0}\leq (q-a_{0})q+a_{1}+1,
				\end{cases}$ $\Rightarrow$
				$\begin{cases}
					a_{1}<a_{0},\\
					a_{1}<\frac{q}{2},\\
					a_{1}+a_{0}<q.
				\end{cases}$
			\end{center}
			Collecting all discussions above, we have $a\in \MinRep_{n}$ if and only if $a\in [l(q+1)+1,(l+1)(q-1)]$, where $0\leq l\leq \frac{q}{2}-1$. 
			
			For any $a\in \MinRep_{n}$, let $|C_{a}|=l$. Note that $ord_{n}(q)=4$, then $l=1,2$ or $4$. If $l=2$, then $aq^{2}\equiv a\pmod {q^{2}+1}$, i.e., $q^{2}+1\mid a(q^{2}-1)$. Since $gcd(q^{2}+1,q^{2}-1)=1$, then $q^{2}+1\mid a$, which is impossible. Similarly, we can get $l\neq 1$. Hence, $|C_{a}|=4$ for any $a\in \MinRep_{n}$. Thus we complete the proof.
		\end{proof}
	\end{lemma}
	Based on the above Lemmas, we can obtain the parameters of $\C_{(n,q,\delta,1)}$
	and $\C_{(n,q,\delta+1,0)}$ for $\delta_{3,n}+1\leq \delta\leq \delta_{1,n}$ as follows.
	
	\begin{theorem}\label{th1}
		Let $n=q^{m}+1$, where $m=4t+2\geq 2$ and $q$ is an even prime power, then the following statements hold:
		\begin{itemize}
			\item[(1)] The LCD BCH code $\C_{(n,q,\delta,1)}$ has parameters
			\begin{center}			
				$\begin{cases}
					[n,5,d\geq \delta_{2,n}],& \text{if}\ \delta_{2,n}+1\leq \delta\leq \delta_{1,n};\\
					[n,2m+5,d\geq \delta_{3,n}],& \text{if}\ \delta_{3,n}+1\leq \delta\leq \delta_{2,n}.
				\end{cases}$	
			\end{center}
			\item[(2)] The LCD BCH code $\C_{(n,q,\delta+1,0)}$ has parameters
			\begin{center}
				$\begin{cases}
					[n,4,d\geq 2\delta_{2,n}],& \text{if}\ \delta_{2,n}+1\leq \delta\leq \delta_{1,n};\\
					[n,2m+4,d\geq 2\delta_{3,n}],& \text{if}\ \delta_{3,n}+1\leq \delta\leq \delta_{2,n}.
				\end{cases}$
			\end{center}
		\end{itemize}
		\begin{proof}
			For any $a\in C_{i} $, it is easy to check that $n-a\in C_{i}$. From the BCH bound, we have $d(\C_{(n,q,\delta,1)})\geq \delta$ and $d(\C_{(n,q,\delta+1,0)})\geq 2\delta$. The dimensions of $\C_{(n,q,\delta,1)}$ and $\C_{(n,q,\delta+1,0)}$ can be obtained directly from Lemmas \ref{le3}, \ref{le7} and \ref{le8}.
		\end{proof}
	\end{theorem}
	\begin{example}
		We provide the following examples based on Theorem \ref{th1}.
		\\$\bullet$ Let $(q,m)=(4,2)$, we have $n=17$ and $\delta_{1,n}=6$. Then LCD BCH code $\C_{(n,q,\delta_{1,n}+1,0)}$ has parameters $[17,4,12]$, the code is distance-optimal according to the tables of best codes known in (\cite{RefJ10}).
		\\$\bullet$ Let $(q,m)=(8,2)$, we have $n=65$ and $\delta_{1,n}=28$. Then LCD BCH code $\C_{(n,q,\delta_{1,n}+1,0)}$ has parameters $[65,4,56]$, the code is distance-optimal according to the tables of best codes known in (\cite{RefJ10}).
	\end{example}
	
	\subsection{When $q$ is an odd prime power}
	
	\begin{lemma}(\cite{RefJ21},\cite{RefJ24})\label{le9}
		Let $n=\frac{q^{m}+1}{\lambda}$ and $q$ be an odd prime power, where $m\geq 2$. Then 
		\begin{itemize}
			\item[(a)] If $\lambda=1$ and $\ q^{m}\equiv1\pmod{4}$, we have the first two largest odd coset leaders modulo $n$ are  $$\delta_{1}=\frac{q^{m}+1}{2},\quad \delta_{2}=\frac{q^{m}-1}{2}-q^{m-1}.$$
			Moreover, $|C_{\delta_{1}}|=1$ and $|C_{\delta_{2}}|=2m$.
			\item[(b)] If $\lambda=2$ and $m=4t+2\geq 6$, we have  $$\delta_{1,n}=\frac{(q-1)^{2}(q^{m}+1)}{4(q^{2}+1)},\quad \delta_{2,n}=\frac{(q-1)^{2}(q^{m}-2q^{m-4}-1)}{4(q^{2}+1)}.$$
			Moreover, $|C_{\delta_{1,n}}|=4$ and $|C_{\delta_{2,n}}|=2m$. 
		\end{itemize}	
	\end{lemma}
	\begin{lemma}\label{le10}
		Let $q$ be an odd prime power and $m\equiv2 \pmod{4}$ with $m\geq 6$, then 
		\begin{center}
			$\delta_{3}=\frac{(q-1)^{2}(q^{m}-2q^{m-4}-1)}{2(q^{2}+1)}$ and 
			$\delta_{4}=$
			$\begin{cases}
				\delta_{3}-q(q-1)^{2}, &\text{if}\ m=6;\\
				\delta_{3}-(q-1)^{2}(q^{2}-1), &\text{if}\ m>6.	
			\end{cases}$
		\end{center}
		are the third and fourth largest coset leaders modulo $n$. Moreover, $|C_{\delta_{3}}|=|C_{\delta_{4}}|=2m$.
		\begin{proof}
			We prove this Lemma in two steps.
			
			$\bf{Step 1}$. We first prove that $\delta_{3,n}=\delta_{3}$. From Lemmas \ref{le2} and \ref{le9}, we have $\delta_{3}=2\delta_{2,\frac{n}{2}}\in \MinRep_{n}$ and $|C_{\delta_{3}}|=|C_{\delta_{2,\frac{n}{2}}}|=2m$, where $C_{\delta_{2,\frac{n}{2}}}$ is the $q$-cyclotomic coset modulo $\frac{n}{2}$ of representative $\delta_{2,\frac{n}{2}}$.
			
			Next we claim that $\delta_{3}$ is the third largest coset leader. Suppose there exists an integer $a$ such that $a\in \MinRep_{n}$ and $\delta_{3}<a<\delta_{2,n}$, where $\delta_{2,n}=\frac{(q-1)^{2}(q^{m}+1)}{2(q^{2}+1)}$ by Lemma \ref{le3}.
			\begin{itemize}
				\item[$\bf{(1)}$] Note that $q$ is an odd prime power, then $q^{m}\equiv3\pmod4$ or $q^{m}\equiv1\pmod4$, but $q^{m}\equiv3\pmod4$ if and only if $q\equiv3\pmod4$ and $2\nmid m$, then $q^{m}\equiv1 \pmod4$. Hence, $\frac{q^{m}+1}{2}$ and $\frac{q^{m}-1}{2}-q^{m-1}$ are the first two largest odd coset leaders by Lemma \ref{le9}. Since  $\frac{q^{m}-1}{2}-q^{m-1}<\delta_{3}<a<\frac{q^{m}+1}{2}$ and $a\in \MinRep_{n}$, then $a$ is not odd.
				\item[$\bf{(2)}$] If $2\mid a$, then $\frac{a}{2}\in \MinRep_{\frac{q^{m}+1}{2}}$ and $\frac{\delta_{3}}{2}<\frac{a}{2}<\frac{\delta_{2,n}}{2}$ by Lemma \ref{le2}. Note that  $\delta_{2,\frac{q^{m}+1}{2}}=\frac{\delta_{3}}{2}$ and $\delta_{1,\frac{q^{m}+1}{2}}=\frac{\delta_{2,n}}{2}$ by Lemma \ref{le9}, which contradicts to $\frac{a}{2}\in \MinRep_{\frac{q^{m}+1}{2}}$.
			\end{itemize}
			Therefore, $\delta_{3}$ is the third largest coset leader.
			
			$\bf{Step 2}$. We prove that $\delta_{4,n}=\delta_{4}$.
			\\$\bf{Case\ 1}$. When $m>6$, it is clear that
			$$\left\{\begin{aligned}
				&\delta_{3}=(\frac{q-3}{2},\frac{q-1}{2},\frac{q+1}{2},\frac{q-3}{2},\underbrace{\underbrace{\frac{q+1}{2},\frac{q-1}{2},\frac{q-3}{2},\frac{q-1}{2}},\ldots}_{\frac{m-6}{4}},\frac{q+1}{2},\frac{q-1}{2}),\\
				&\delta_{4}=(\frac{q-3}{2},\frac{q-1}{2},\frac{q+1}{2},\frac{q-3}{2},\underbrace{\underbrace{\frac{q+1}{2},\frac{q-1}{2},\frac{q-3}{2},\frac{q-1}{2}},\ldots}_{\frac{m-10}{4}},\frac{q+1}{2},\frac{q-3}{2},\frac{q+1}{2},\frac{q-1}{2},\frac{q-3}{2},\frac{q+1}{2}).\end{aligned}
			\right.$$
			From the Lemma \ref{le1}, we can get $\delta_{4}\in \MinRep_{n}$ and $|C_{\delta_{4}}|=2m$ by the same way of Lemma \ref{le7}. 
			
			We assume that $a\in \MinRep_{n}$ and $\delta_{4}<a<\delta_{3}$, then the $q$-adic expansion of $a$ can be written as $$a=(\frac{q-3}{2},\frac{q-1}{2},\frac{q+1}{2},\frac{q-3}{2},\underbrace{\underbrace{\frac{q+1}{2},\frac{q-1}{2},\frac{q-3}{2},\frac{q-1}{2}},\ldots}_{\frac{m-10}{4}},\frac{q+1}{2},a_{4},a_{3},a_{2},a_{1},a_{0}),$$
			where $a_{4}\leq \frac{q-1}{2}$ and $a_{0}\neq 0$. Similar to the proof of Lemma \ref{le7}, we have $a_{i}\in \{\frac{q-3}{2}, \frac{q-1}{2}, \frac{q+1}{2}\}$ for $0\leq i\leq 4$. Since $a\leq [aq^{m-1}]_{n}$, then $ \frac{q-1}{2}\leq a_{0}$.
			\begin{itemize}
				\item[$\bf{(1)}$] If $a_{4}=\frac{q-1}{2}$. Note that $a<\delta_{3,n}$, then we have $a_{3}=\frac{q-3}{2}$, $a_{2}=\frac{q-1}{2}$ or $\frac{q-3}{2}$, and it must have $a_{1}=\frac{q-1}{2}$ or $\frac{q-3}{2}$ when $a_{2}=\frac{q-1}{2}$. Clearly, we have $$[aq^{m-4}]_{n}\leq (\frac{q-3}{2},\frac{q-1}{2},\frac{q-1}{2},\underbrace{q-1,\ldots}_{m-4},q-1)<a.$$
				\item[$\bf{(2)}$]If $a_{4}=\frac{q-3}{2}$. Note that $\delta_{4,n}<a$, then we have $a_{3}=\frac{q+1}{2}$, $a_{2}=\frac{q+1}{2}$ or $\frac{q-1}{2}$, and it must have $a_{1}=\frac{q+1}{2}$ or $\frac{q-1}{2}$ when $a_{2}=\frac{q-1}{2}$.
				Clearly, we have $$n-[aq^{m-4}]_{n}\leq (\frac{q-3}{2},\frac{q-1}{2},\frac{q-1}{2},\underbrace{q-1,\ldots}_{m-4},q-1)<a.$$
			\end{itemize}
			Hence, $\delta_{4}$ is the fourth largest coset leader.
			\\$\bf{Case\ 2}$. When $m=6$, it is clear that $\delta_{4}\leq [\delta_{4}q^{i}]_{n}$, $n-[\delta_{4}q^{i}]_{n}$ for any $0\leq i\leq 5$, then $\delta_{4}\in \MinRep_{n}$ and $|C_{\delta_{4}}|=2m$.
			Similar to $\bf{Case\ 1}$, we can get $\delta_{4,n}=\delta_{4}$.
			
			Thus we complete the proof.
		\end{proof}
	\end{lemma}
	
	\begin{lemma}\label{le11}
		Let $n=q^{2}+1$ and $q$ be an odd prime power. Then $a\in \MinRep_{n}$ if and only if $a\in [l(q+1)+1,(l+1)(q-1)]$ or $a=\frac{q^{2}+1}{2}$, where $0\leq l\leq \frac{q-3}{2}$. Moreover, $|C_{\frac{q^{2}+1}{2}}|=1$ and $|C_{a}|=4$ for any $a\in \MinRep_{n}\textbackslash \{\frac{q^{2}+1}{2}\}$.
		\begin{proof}
			The proof is similar to Lemma \ref{le8}, hence we omit the proof.
		\end{proof}
	\end{lemma}
	Based on the above Lemmas, we can obtain the parameters of $\C_{(n,q,\delta,1)}$
	and $\C_{(n,q,\delta+1,0)}$ for $\delta_{4,n}+1\leq \delta\leq \delta_{2,n}$ as follows.
	\begin{theorem}\label{th2}
		Let $n=q^{m}+1$, where $m=4t+2\geq 2$ and $q$ is an odd prime power, then the following statements hold:
		\begin{itemize}\label{t18}
			\item[(1)] The LCD BCH code $C_{(n,q,\delta,1)}$ has parameters
			\begin{center}			
				$\begin{cases}
					[n,6,d\geq \delta_{3,n}],& \text{if}\ \delta_{3,n}+1\leq \delta\leq \delta_{2,n};\\
					[n,2m+6,d\geq \delta_{4,n}],& \text{if}\ \delta_{4,n}+1\leq \delta\leq \delta_{3,n}.
				\end{cases}$	
			\end{center}
			\item[(2)] The LCD BCH code $C_{(n,q,\delta+1,0)}$ has parameters
			\begin{center}
				$\begin{cases}
					[n,5,d\geq 2\delta_{3,n}],& \text{if}\ \delta_{3,n}+1\leq \delta\leq \delta_{2,n};\\
					[n,2m+5,d\geq 2\delta_{4,n}],& \text{if}\ \delta_{4,n}+1\leq \delta\leq \delta_{3,n}.
				\end{cases}$
			\end{center}
		\end{itemize}
		\begin{proof}
			The results can be obtained directly by Lemmas \ref{le3}, \ref{le10} and \ref{le11}, hence we omit the details.
		\end{proof}
	\end{theorem}
	\begin{example}
		We provide the following examples based on Theorem \ref{th2}.
		\\$\bullet$ Let $(q,m)=(7,2)$, we have $n=50$ and $\delta_{3,n}=17$. Then LCD BCH code $\C_{(n,q,\delta_{3,n}+1,0)}$ has parameters $[50,9,34]$, the code is distance-optimal according to the tables of best codes known in (\cite{RefJ10}).
		\\$\bullet$ Let $(q,m)=(9,2)$, we have $n=82$ and $\delta_{3,n}=31$. Then LCD BCH code $\C_{(n,q,\delta_{3,n}+1,0)}$ has parameters $[82,9,62]$, the code is distance-optimal according to the tables of best codes known in (\cite{RefJ10}).
	\end{example}
	\section{The LCD BCH Codes of Length $n=\frac{q^{m}+1}{\lambda}$ with $\lambda\neq1$}
	
	Throughout this section, let $n=\frac{q^{m}+1}{\lambda}$ and $\lambda\neq 1$ be a positive divisor of $q+1$, where $q$ is an odd prime power.
	\subsection{The case of $q+1>\lambda\geq 2$}
	In this subsection, we discuss the coset leaders based on the value of $m$. We first consider the case of $m=2t+1\geq 5$, where $t$ is an integer.
	\begin{lemma}\label{le12}
		Let $n=\frac{q^{m}+1}{\lambda}$, where $m=2t+1\geq 5$. Let $f(a,b,c)=\frac{q^{t+1}+1}{\lambda}+a\frac{q^{t}-1}{\lambda}+bq^{t}+c$ and $g(a,b,c,d)=\frac{q^{t+1}-1}{\lambda}+a\frac{q^{t}+1}{\lambda}+bq^{t}+c$, where $a,b,c,d$ are integers.	We assume that
		\begin{center}
			If $2\mid t$,
			$\begin{cases}
				T_{1}=\lbrace f(0,0,c): -\frac{q}{\lambda} \leq c\leq \frac{q-2}{\lambda}  \rbrace;\\
				T_{2}=\lbrace f(0,b,0):1\leq b\leq \frac{q-1}{\lambda} \rbrace;\\	
				T_{3}=\lbrace f(2,b,0):0 \leq b\leq \frac{q-3}{\lambda}\rbrace.
			\end{cases}$If $2\nmid t$, $\begin{cases}
				T_{1}=\lbrace g(0,0,c):-\frac{q-2}{\lambda}\leq c\leq \frac{q}{\lambda}  \rbrace;\\		
				T_{2}=\lbrace g(0,b,0):1\leq b\leq \frac{q-1}{\lambda}  \rbrace;\\
				T_{3}=\lbrace g(2,b,0):0 \leq b\leq \frac{q-3}{\lambda} \rbrace.
			\end{cases}$
		\end{center} Then we have
		\begin{itemize} 
			\item [(1)]
			If $1\leq i< \frac{2q^{t+1}-2q+1}{\lambda}$ with $i\not\equiv0\pmod{q}$, then $i$ is a coset leader modulo $\frac{q^{m}+1}{\lambda}$ except $i\in T_{1}\cup T_{2}\cup T_{3}$ and $|C_{i}| =2m$.
			\item[(2)]
			If $\frac{2q^{t+1}-2q+1}{\lambda}\leq i\leq \frac{2q^{t+1}+2q-1}{\lambda}$, then $i$ is not a coset leader.
		\end{itemize}
		
		\begin{proof}
			We only prove the conclusion (1), as the proof of (2) is similar.
			If $1\leq i< \frac{2q^{t+1}-2q+1}{\lambda}$ with $i\not\equiv0\pmod{q}$, we know that $i\notin\MinRep_{n}$ if and only if $\lambda i\notin\MinRep_{q^{m}+1}$ with $\lambda i\not\equiv0\pmod{q}$ by Lemma \ref{le2}. From Lemma \ref{le4}, we know that $\lambda i\notin\MinRep_{q^{m}+1}$ with $\lambda i\not\equiv0\pmod{q}$ only when 
			\begin{center}
				$\lambda i= q^{t+1}\pm \gamma$	or $\lambda i=q^{t+1}+\alpha q^{t}\pm 1$,
			\end{center}
			where $\gamma, \alpha\in [1,q-1]$.
			
			We divide $t$ into two cases to get our results.
			\begin{itemize}
				\item [$\bf{(i)}$] If $2\mid t$, then $\lambda\mid (q^{t+1}+1)$ and $\lambda\mid(q^{t}-1)$.
				
				For $\lambda i=q^{t+1}\pm \gamma$ with $1\leq \gamma\leq q-1$, then $i=\frac{q^{t+1}+1}{\lambda}+\frac{l}{\lambda}$ with $l\in [-q,-2]\cup[0,q-2]$. Since $\lambda\geq 2$, then $\frac{-1}{\lambda}$ is not an integer, i.e., $\{i=\frac{q^{t+1}+1}{\lambda}+c:-\frac{q}{\lambda}\leq c\leq \frac{q-2}{\lambda}\}=T_{1}$.
				
				For $\lambda i=q^{t+1}+\alpha q^{t}+ 1$, we get that $i=\frac{q^{t+1}+1}{\lambda}+\frac{\alpha}{\lambda}q^{t}$. Note that $1\leq \alpha\leq q-1$, then we have $\frac{1}{\lambda}\leq \frac{\alpha}{\lambda}\leq \frac{q-1}{\lambda}$. Since $\lambda\geq 2$, then $\{i=\frac{q^{t+1}+1}{\lambda}+bq^{t}:1\leq b\leq \frac{q-1}{\lambda}\}= T_{2}$.
				
				For $\lambda i=q^{t+1}+\alpha q^{t}-1=(q^{t+1}+1)+(\alpha-2)q^{t}+2(q^{t}-1)$, we get that $i=\frac{q^{t+1}+1}{\lambda}+\frac{2(q^{t}-1)}{\lambda}+\frac{\alpha-2}{\lambda}q^{t}$. Since $\lambda\geq 2$ and $1\leq \alpha\leq q-1$, then $\{i=\frac{q^{t+1}+1}{\lambda}+\frac{2(q^{t}-1)}{\lambda}+bq^{t}:0\leq b\leq \frac{q-3}{\lambda}\}= T_{3}$.
				\item [$\bf{(ii)}$] If $2\nmid t$, then $\lambda\mid (q^{t}+1)$ and $\lambda\mid (q^{t+1}-1)$. With the same discussion of the case $2\mid t$, we can obtain $T_{1},T_{2}$ and $T_{3}$.
			\end{itemize}
			
			Moreover, we have $|C_{i}| =|C_{\lambda i}|=2m$ by Lemmas \ref{le2} and \ref{le4}. Thus we complete the proof.
		\end{proof}
	\end{lemma}
	\begin{corollary}\label{co1}
		Let $n=\frac{q^{m}+1}{\lambda}$ and $m=2t+1\geq 5$, then
		\begin{itemize}
			\item[(1)]If $2\mid t$, the smallest $i$ such that $i\not\equiv 0\pmod q$ and $i\notin\MinRep_{n}$ is $\frac{q^{t+1}+1}{\lambda}-\lfloor \frac{q}{\lambda}\rfloor$.
			\item[(2)]If $2\nmid t$, the smallest $i$ such that $i\not\equiv 0\pmod q$ and $i\notin\MinRep_{n}$ is $\frac{q^{t+1}-1}{\lambda}-\lfloor \frac{q-2}{\lambda}\rfloor$.
		\end{itemize}
		\begin{proof}
			Note that $t\geq 2$ and $2\leq \lambda\leq \frac{q+1}{2}$, it is clear that $min(T_{1},T_{2},T_{3})=$
			$\begin{cases}
				\frac{q^{t+1}+1}{\lambda}-\lfloor \frac{q}{\lambda}\rfloor, &\ if \ 2\mid t\\
				\frac{q^{t+1}-1}{\lambda}-\lfloor \frac{q-2}{\lambda}\rfloor, &\ if \ 2\nmid t
			\end{cases}$, where the definition of $T_{1}, T_{2},T_{3}$ are given by Lemma \ref{le12}. Thus we complete the proof.
		\end{proof}
	\end{corollary}
	Next we will investigate the parameters of the BCH codes $\C_{(n,q,\delta,1)}$ and $\C_{(n,q,\delta+1,0)}$, where $1\leq \delta-1\leq \frac{2q^{t+1}+2q-1}{\lambda} $.
	
	\begin{theorem}\label{th3}
		Let $n=\frac{q^{m}+1}{\lambda}$ and $\epsilon=\lceil (\delta-1)(1-q^{-1})\rceil$, where $m=2t+1\geq 5$ and $2\leq \lambda<q+1$.
		If $2\mid t$ and $1\leq \delta-1\leq \frac{2q^{t+1}+2q-1}{\lambda}$, then LCD BCH code $\C_{(n,q,\delta,1)}$ has parameters $[\frac{q^{m}+1}{\lambda},k,d\geq \delta]$ and $\C_{(n,q,\delta+1,0)}$ has parameters $[\frac{q^{m}+1}{\lambda},k-1,d\geq 2\delta]$, where $k$ is given as follows.
		\\(1) If $2\leq \delta\leq \frac{q^{t+1}+1}{\lambda}+ \frac{2(q^{t}-1)}{\lambda}$, define $\upsilon=\lfloor\frac{q-2}{\lambda}\rfloor+\lfloor\frac{q}{\lambda} \rfloor$. Then
		\begin{center}
			$k=$$\begin{cases}
				n-2m\epsilon,& \text{if}\ 2\leq \delta\leq \frac{q^{t+1}+1}{\lambda}-\lfloor\frac{q}{\lambda} \rfloor;\\
				n-2m\epsilon+2m(\delta-\frac{q^{t+1}+1}{\lambda}+\lfloor\frac{q}{\lambda} \rfloor),& \text{if}\ \frac{q^{t+1}+1}{\lambda}-\lfloor\frac{q}{\lambda} \rfloor<\delta\leq \frac{q^{t+1}+1}{\lambda}+\lfloor\frac{q-2}{\lambda}\rfloor;\\
				n-2m(\epsilon-\upsilon-1),& \text{if}\ \frac{q^{t+1}+1}{\lambda}+\lfloor\frac{q-2}{\lambda}\rfloor<\delta\leq \frac{q^{t+1}+1}{\lambda}+\frac{2(q^{t}-1)}{\lambda}.
			\end{cases}$
		\end{center}
		(2) If $\left\lfloor \frac{q-3}{\lambda} \right\rfloor\neq 0$ and $\frac{q^{t+1}+1}{\lambda}+ \frac{2(q^{t}-1)}{\lambda}<\delta\leq \frac{q^{t+1}+1}{\lambda}+\frac{2(q^{t}-1)}{\lambda}+\left\lfloor \frac{q-3}{\lambda} \right\rfloor q^{t} $, let $a\in [0,\lfloor \frac{q-3}{\lambda}\rfloor-1]$ be an integer. Then 
		\begin{center}
			$k=$$\begin{cases}
				n-2m(\epsilon-\upsilon-2a-2),& \text{if}\ \frac{q^{t+1}+1}{\lambda}+\frac{2(q^{t}-1)}{\lambda}+aq^{t}<\delta\leq \frac{q^{t+1}+1}{\lambda}+(a+1)q^{t};\\
				n-2m(\epsilon-\upsilon-2a-3),& \text{if}\ \frac{q^{t+1}+1}{\lambda}+(a+1)q^{t}<\delta\leq \frac{q^{t+1}+1}{\lambda}+\frac{2(q^{t}-1)}{\lambda}+(a+1)q^{t}.
			\end{cases}$
		\end{center}
		(3) If $\frac{q^{t+1}+1}{\lambda}+\frac{2(q^{t}-1)}{\lambda}+\left\lfloor \frac{q-3}{\lambda} \right\rfloor q^{t}<\delta\leq \frac{2q^{t+1}+2q-1}{\lambda}+1$, define $\kappa=2\frac{q^{t+1}+1}{\lambda}-\lfloor \frac{2q+1}{\lambda}\rfloor$ and $\tau=\lceil (\kappa-1)(1-q^{-1})\rceil$. 
		\begin{itemize}
			\item[(i)]If $\left\lfloor \frac{q-1}{\lambda} \right\rfloor=\left \lfloor \frac{q-3}{\lambda} \right\rfloor$, then 
			\begin{center}
				$k=$$\begin{cases}
					n-2m(\epsilon-\upsilon-2\lfloor \frac{q-1}{\lambda}\rfloor-2),& \text{if}\ \frac{q^{t+1}+1}{\lambda}+\frac{2(q^{t}-1)}{\lambda}+\lfloor \frac{q-3}{\lambda}\rfloor q^{t} <\delta\leq \kappa;\\
					n-2m(\tau-\upsilon-2\lfloor \frac{q-1}{\lambda}\rfloor-2),& \text{if}\ \kappa<\delta\leq \frac{2q^{t+1}+2q-1}{\lambda}+1.
				\end{cases}$
			\end{center}
			\item[(ii)]If $\left\lfloor \frac{q-1}{\lambda} \right\rfloor=\left \lfloor \frac{q-3}{\lambda} \right\rfloor+1$, then 
			$$k=\begin{cases}
				n-2m(\epsilon-\upsilon-2\lfloor \frac{q-1}{\lambda}\rfloor),& \text{if}\ \frac{q^{t+1}+1}{\lambda}+\frac{2(q^{t}-1)}{\lambda}+\lfloor \frac{q-3}{\lambda}\rfloor q^{t}<\delta\leq \frac{q^{t+1}+1}{\lambda}+\lfloor \frac{q-1}{\lambda}\rfloor q^{t};\\
				n-2m(\epsilon-\upsilon-2\lfloor \frac{q-1}{\lambda}\rfloor-1),& \text{if}\ \frac{q^{t+1}+1}{\lambda}+\lfloor \frac{q-1}{\lambda}\rfloor q^{t} <\delta\leq \kappa;\\
				n-2m(\tau-\upsilon-2\lfloor \frac{q-1}{\lambda}\rfloor-1),& \text{if}\ \kappa<\delta\leq \frac{2q^{t+1}+2q-1}{\lambda}+1.
			\end{cases}$$
		\end{itemize} 
		\begin{proof}
			We only prove the conclusion (3), as the proofs for other cases are similar.
			It is clear that $d(\C_{(n,q,\delta,1)})\geq \delta$ and $d(\C_{(n,q,\delta+1,0)})\geq 2\delta$ by the BCH bound. $T$ and $T^{'}$ are denoted as the defining sets of $\C_{(n,q,\delta,1)}$ and $\C_{(n,q,\delta+1,0)}$, respectively. Note that $T=T^{'}\textbackslash \{0\}$, we have $dim(\C_{(n,q,\delta,1)})=dim(\C_{(n,q,\delta+1,0)})+1$, then we only need to discuss the dimension of $\C_{(n,q,\delta,1)}$.
			\\$\bf{Case\ 1}$: For $\frac{q^{t+1}+1}{\lambda}+\frac{2(q^{t}-1)}{\lambda}+\left\lfloor \frac{q-3}{\lambda} \right\rfloor q^{t}<\delta\leq\kappa$, let $\Gamma=\lbrace a:1\leq a\leq \delta-1,\ \ a\not\equiv 0\pmod q\rbrace$, then $i\in \Gamma$ is a coset leader except $i\in T_{1}\cup T_{2}\cup T_{3}$ and $|C_{i}|=2m$ for any $\delta\leq \lceil \frac{2q^{t+1}-2q+1}{\lambda}\rceil=\kappa$ by Lemma \ref{le12},  where $T_{1}, T_{2},T_{3}$ are given by Lemma \ref{le12}. Then
			\begin{equation}\label{eq:1}
				k=n-2m(\mid \Gamma \mid-\sum_{i=1}^{3}\mid \Gamma\cap T_{i}\mid),
			\end{equation}
			it is clear that $\mid \Gamma \mid =\lceil (\delta-1)(1-q^{-1})\rceil$, then we need to calculate the values of $\mid \Gamma\cap T_{i} \mid$.
			\begin{itemize}
				\item[$\bf{(i)}$] If $\left\lfloor \frac{q-1}{\lambda} \right\rfloor=\left \lfloor \frac{q-3}{\lambda} \right\rfloor$, we have $max\lbrace T_{1},T_{2},T_{3}\rbrace=\frac{q^{t+1}+1}{\lambda}+\frac{2(q^{t}-1)}{\lambda}+\lfloor \frac{q-3}{\lambda}\rfloor q^{t}\leq \delta-1$. Then 
				\begin{center}
					$\Gamma\cap T_{1}=\lbrace \frac{q^{t+1}+1}{\lambda}+c: -\lfloor\frac{q}{\lambda} \rfloor \leq c \leq \lfloor\frac{q-2}{\lambda} \rfloor \rbrace$,\\
					$\Gamma\cap T_{2}=\lbrace \frac{q^{t+1}+1}{\lambda}+bq^{t}: 1\leq b\leq \lfloor\frac{q-1}{\lambda} \rfloor \rbrace$,\\
					$\Gamma\cap T_{3}=\lbrace \frac{q^{t+1}+1}{\lambda}+\frac{2(q^{t}-1)}{\lambda}+bq^{t}: 0\leq b\leq \lfloor\frac{q-3}{\lambda} \rfloor \rbrace$.\\
				\end{center}
				It is clear that $\mid \Gamma\cap T_{1}\mid=\lfloor\frac{q-2}{\lambda}\rfloor+\lfloor\frac{q}{\lambda} \rfloor+1=\upsilon+1$, $\mid \Gamma\cap T_{2}\mid=\lfloor\frac{q-1}{\lambda}\rfloor$ and  $\mid \Gamma\cap T_{3}\mid=\lfloor\frac{q-1}{\lambda}\rfloor+1$, then $dim(\C_{(n,q,\delta,1)})$ can be obtained by (\ref{eq:1}). 
				\item[$\bf{(ii)}$] If $\left\lfloor \frac{q-1}{\lambda} \right\rfloor=\left \lfloor \frac{q-3}{\lambda} \right\rfloor+1$, note that $i\in T_{1}\cup T_{2}\cup T_{3}$ and $\frac{q^{t+1}+1}{\lambda}+\frac{2(q^{t}-1)}{\lambda}+\lfloor \frac{q-3}{\lambda}\rfloor q^{t}<i$ if and only if $i=\frac{q^{t+1}+1}{\lambda}+\lfloor \frac{q-1}{\lambda}\rfloor q^{t}\in T_{2}$. Then we have
				\begin{center}
					$\Gamma\cap T_{1}=\lbrace \frac{q^{t+1}+1}{\lambda}+c: -\lfloor\frac{q}{\lambda} \rfloor \leq c \leq \lfloor\frac{q-2}{\lambda} \rfloor \rbrace$,\\
					$\Gamma\cap T_{3}=\lbrace \frac{q^{t+1}+1}{\lambda}+\frac{2(q^{t}-1)}{\lambda}+bq^{t}: 0\leq b\leq \lfloor\frac{q-3}{\lambda} \rfloor \rbrace$,\\	
					$\Gamma\cap T_{2}=$	
					$\begin{cases}
						\lbrace \frac{q^{t+1}+1}{\lambda}+bq^{t}: 1\leq b\leq \lfloor\frac{q-3}{\lambda} \rfloor \rbrace,& \text{if}\ \omega <\delta\leq \frac{q^{t+1}+1}{\lambda}+\lfloor \frac{q-1}{\lambda}\rfloor q^{t};\\
						\lbrace \frac{q^{t+1}+1}{\lambda}+bq^{t}: 1\leq b\leq \lfloor\frac{q-1}{\lambda} \rfloor \rbrace,& \text{if}\ \frac{q^{t+1}+1}{\lambda}+\lfloor \frac{q-1}{\lambda}\rfloor q^{t}<\delta\leq \kappa,
					\end{cases}$
				\end{center}
				where $\omega=\frac{q^{t+1}+1}{\lambda}+\frac{2(q^{t}-1)}{\lambda}+\lfloor \frac{q-3}{\lambda}\rfloor q^{t}$, 
				then $dim(\C_{(n,q,\delta,1)})$ can be obtaioned by (\ref{eq:1}). 
			\end{itemize}
			$\bf{Case\ 2}$: For $\kappa<\delta\leq \frac{2q^{t+1}+2q-1}{\lambda}+1$, we have $i\notin \MinRep_{n}$ for any $i\in [\kappa,\delta-1]$, then $dim(\C_{(n,q,\delta,1)})=dim(\C_{(n,q,\kappa,1)})$.
			\\Thus we complete the proof.
		\end{proof}
	\end{theorem}
	
	\begin{theorem}\label{th4}
		Let $n=\frac{q^{m}+1}{\lambda}$ and $\epsilon=\lceil (\delta-1)(1-q^{-1})\rceil$, where $m=2t+1\geq 5$ and $2\leq \lambda<q+1$.
		If $2\nmid t$ and $1\leq \delta-1\leq \frac{2q^{t+1}+2q-1}{\lambda}$, then LCD BCH code $\C_{(n,q,\delta,1)}$ has parameters $[\frac{q^{m}+1}{\lambda},k,d\geq \delta]$ and $\C_{(n,q,\delta+1,0)}$ has parameters $[\frac{q^{m}+1}{\lambda},k-1,d\geq 2\delta]$, where $k$ is given as follows.
		\\(1) If $2\leq \delta\leq \frac{q^{t+1}-1}{\lambda}+ \frac{2(q^{t}+1)}{\lambda}$, define $\upsilon=\lfloor\frac{q-2}{\lambda}\rfloor+\lfloor\frac{q}{\lambda} \rfloor$. Then 
		\begin{center}
			$k=$$\begin{cases}
				n-2m\epsilon,& \text{if}\ 2\leq \delta\leq \frac{q^{t+1}-1}{\lambda}-\lfloor\frac{q-2}{\lambda} \rfloor;\\
				n-2m\epsilon+2m(\delta-\frac{q^{t+1}-1}{\lambda}+\lfloor\frac{q-2}{\lambda} \rfloor),& \text{if}\ \frac{q^{t+1}-1}{\lambda}-\lfloor\frac{q-2}{\lambda} \rfloor<\delta\leq \frac{q^{t+1}-1}{\lambda}+\lfloor\frac{q}{\lambda}\rfloor;\\
				n-2m(\epsilon-\upsilon-1),& \text{if}\ \frac{q^{t+1}-1}{\lambda}+\lfloor\frac{q}{\lambda}\rfloor<\delta\leq \frac{q^{t+1}-1}{\lambda}+\frac{2(q^{t}+1)}{\lambda}.
			\end{cases}$
		\end{center}
		(2) If $\left\lfloor \frac{q-3}{\lambda} \right\rfloor\neq 0$ and  $\frac{q^{t+1}-1}{\lambda}+ \frac{2(q^{t}+1)}{\lambda}<\delta\leq \frac{q^{t+1}-1}{\lambda}+\frac{2(q^{t}+1)}{\lambda}+\left\lfloor \frac{q-3}{\lambda} \right\rfloor q^{t} $, let $a\in [0,\lfloor \frac{q-3}{\lambda}\rfloor-1]$ be an integer. Then
		\begin{center}
			$k=$$\begin{cases}
				n-2m(\epsilon-\upsilon-2a-2),& \text{if}\ \frac{q^{t+1}-1}{\lambda}+\frac{2(q^{t}+1)}{\lambda}+aq^{t}<\delta\leq \frac{q^{t+1}-1}{\lambda}+(a+1)q^{t};\\
				n-2m(\epsilon-\upsilon-2a-3),& \text{if}\ \frac{q^{t+1}-1}{\lambda}+(a+1)q^{t}<\delta\leq \frac{q^{t+1}-1}{\lambda}+\frac{2(q^{t}+1)}{\lambda}+(a+1)q^{t}.
			\end{cases}$
		\end{center}
		(3) If $\frac{q^{t+1}-1}{\lambda}+\frac{2(q^{t}+1)}{\lambda}+\left\lfloor \frac{q-3}{\lambda} \right\rfloor q^{t}<\delta\leq \frac{2q^{t+1}+2q-1}{\lambda}+1$, define $\kappa=2\frac{q^{t+1}-1}{\lambda}-\lfloor \frac{2q-3}{\lambda}\rfloor$ and $\tau=\lceil (\kappa-1)(1-q^{-1})\rceil$. 
		\begin{itemize}		
			\item[(i)]If $\left\lfloor \frac{q-1}{\lambda} \right\rfloor=\left \lfloor \frac{q-3}{\lambda} \right\rfloor$, then 
			$$k=\begin{cases}
				n-2m(\epsilon-\upsilon-2\lfloor \frac{q-1}{\lambda}\rfloor-2),& \text{if}\ \frac{q^{t+1}-1}{\lambda}+\frac{2(q^{t}+1)}{\lambda}+\lfloor \frac{q-3}{\lambda}\rfloor q^{t} <\delta\leq \kappa;\\
				n-2m(\tau-\upsilon-2\lfloor \frac{q-1}{\lambda}\rfloor-2),& \text{if}\ \kappa<\delta\leq \frac{2q^{t+1}+2q-1}{\lambda}+1.
			\end{cases}$$
			\item[(ii)]If $\left\lfloor \frac{q-1}{\lambda} \right\rfloor=\left \lfloor \frac{q-3}{\lambda} \right\rfloor+1$, then 
			$$k=\begin{cases}
				n-2m(\epsilon-\upsilon-2\lfloor \frac{q-1}{\lambda}\rfloor),& \text{if}\ \frac{q^{t+1}-1}{\lambda}+\frac{2(q^{t}+1)}{\lambda}+\lfloor \frac{q-3}{\lambda}\rfloor q^{t}<\delta\leq \frac{q^{t+1}-1}{\lambda}+\lfloor \frac{q-1}{\lambda}\rfloor q^{t};\\
				n-2m(\epsilon-\upsilon-2\lfloor \frac{q-1}{\lambda}\rfloor-1),& \text{if}\ \frac{q^{t+1}-1}{\lambda}+\lfloor \frac{q-1}{\lambda}\rfloor q^{t} <\delta\leq \kappa;\\
				n-2m(\tau-\upsilon-2\lfloor \frac{q-1}{\lambda}\rfloor-1),& \text{if}\ \kappa<\delta\leq \frac{2q^{t+1}+2q-1}{\lambda}+1.
			\end{cases}$$
		\end{itemize}
		\begin{proof}
			From Lemma \ref{le12}, we can obtain the results following the same way of Theorem \ref{th3}.
		\end{proof}
	\end{theorem}
	
	\begin{theorem}\label{th6}
		Let $n=\frac{q^{3}+1}{2}$ and $\epsilon=\lceil (\delta-1)(1-q^{-1})\rceil$. For any integer $\delta$ with $1\leq \delta-1\leq \frac{q^{2}-1}{2}$, then $\C_{(n,q,\delta,1)}$ has parameters $[\frac{q^{3}+1}{2},k,d\geq \delta]$ and $\C_{(n,q,\delta+1,0)}$ has parameters $[\frac{q^{3}+1}{2},k-1,d\geq 2\delta]$, where \begin{center}
			$k=$$\begin{cases}
				n-6\epsilon,& \text{if}\ 2\leq \delta\leq \frac{q^2-1}{2}-\frac{q-3}{2};\\
				n-6\epsilon+6(\delta-\frac{q^2-1}{2}+\lfloor\frac{q-2}{2}\rfloor),& \text{if}\ \frac{q^2-1}{2}-\frac{q-3}{2}<\delta\leq\frac{q^2-1}{2}+1.
			\end{cases}$
		\end{center}
		
		\begin{proof}
			If $1\leq i\leq \frac{q^{2}-1}{2}$ and $i\not\equiv0\pmod{q}$, then we have $i\in\MinRep_{n}$ with $|C_{i}|=2m=6$ except that $i=\frac{q^2-1}{2}-a $ by Lemmas \ref{le2} and \ref{le4}, where $0\leq a\leq \lfloor\frac{q-2}{2}\rfloor$ is an integer. Following the same way of Theorem \ref{th3}, we can obtain the results.
		\end{proof}
	\end{theorem}
	\begin{remark}\label{re1}
		For $n=\frac{q^m+1}{2}$, where $q\equiv3\pmod{4}$ and $m=2t+1\geq 5$. We have $$\delta_{1}=\frac{q^{m}+1}{4},\ \delta_{2}=\frac{q^{m}-1}{4}-\frac{q^{m-1}}{2},\ 
		\delta_{3}=\delta_{2}-\frac{q-1}{2}.$$ 
		are the first there largest coset leaders modulo $n$. Moreover, $|C_{\delta_{1}}|=1$ and $|C_{\delta_{2}}|=|C_{\delta_{3}}|=2m$.	
		\begin{proof}
			By Lemmas \ref{le2} and \ref{le3}, we can get $\delta_{1}\in\MinRep_{n}$ with $|C_{\delta_1}|=1$ and there is no coset leader in $[\delta_{1}+1,n-1]$, i.e., $\delta_{1,n}=\delta_1$. When $q\equiv3\pmod{4}$ and $m=2t+1\geq 5$, we have $\frac{q^{m}-1}{2}-q^{m-1}\in \MinRep_{q^{m}+1} \ and$ $\frac{q^{m}-1}{2}-q^{m-1}+2\mu\notin \MinRep_{q^{m}+1}$ for $1\leq \mu\leq \frac{q^{m-1}-1}{2}$ by the same way of [\cite{RefJ21}, Lemma 22], which implies that $\delta_{2}\in\MinRep_{n}$ with $|C_{\delta_2}|=2m$ and there is no coset leader in $[\delta_{2}+1,\delta_{1}-1]$ by Lemma \ref{le2}, i.e., $\delta_{2,n}=\delta_2$. From Lemma \ref{le2} and [\cite{RefJ21}, Lemma 23], we have $\delta_{3,n}=\delta_3$ and $|C_{\delta_3}|=2m$ similarly.
		\end{proof}
	\end{remark}
	\begin{example}
		Based on the above theorems, we have the following examples.
		\\$\bullet$ Let $(q,m)=(5,3)$ and $\lambda=2$, we have $n=\frac{q^{3}+1}{2}=63$. The LCD BCH code $\C_{(n,q,3,0)}$ has parameters $[63,56,4]$, the code is distance-optimal according to the tables of best codes known in (\cite{RefJ10}).
		\\$\bullet$ Let $(q,m)=(5,5)$ and $\lambda=3$, we have $n=\frac{q^5+1}{3}=1042$. The LCD BCH code $\C_{(n,q,51,0)}$ has parameters $[1042,671,\geq 100]$, the dimension is consistent with what was given in  Theorem \ref{th3}.
	\end{example}
	Next we will consider the case of $m=2t\geq 4$, where $t$ is an integer. In this case, we only cosider that $\lambda=2$. For $n=\frac{q^{m}+1}{2}$, we already know all coset leaders for $1\leq i\leq \frac{q^{t}}{2}$ in \cite{RefJ23}.
	\begin{lemma}(\cite{RefJ22})\label{le13}
		Let $n=q^{m}+1$ and $m=2t\geq8$. For $q^{t}\leq a\leq q^{t+1}$ with $a\not\equiv0\pmod{q}$, then $a$ is a coset leader and $|C_{a}|=2m$ except that $a\in A_{1}\cup A_{2}\cup A_{3}$, where 
		\begin{center}
			$\begin{cases}
				A_{1}=\{a_{t}q^{t}+a_{0}:1\leq a_{0}\leq a_{t}\leq q-1\},\\
				A_{2}=\{a_{t}q^{t}-a_{0}:1\leq a_{0}<a_{t}\leq q-1\},\\
				A_{3}=\{a:q^{t+1}-q^{2}+1\leq a\leq q^{t+1},q\nmid a\}.
			\end{cases}$
		\end{center}
	\end{lemma}
	\begin{lemma}\label{le14}
		Let $n=\frac{q^{m}+1}{2}$ and $m=2t\geq 4$. 
		\begin{itemize}
			\item[(1)]If $\frac{q^{t}+1}{2}\leq i\leq q^{t}+1$ and $i\not\equiv0\pmod{q}$. Then $i$ is a coset leader and $|C_{i}|=2m$ except $i=\frac{q^{t}+1}{2}$ or $i=q^{t}+1$.
			\item[(2)]If $m=2t\geq 8$ and $\frac{q^{t}+1}{2}\leq i\leq \frac{q^{t+1}-1}{2}$ with $i\not\equiv0\pmod{q}$.
			Then $i$ is a coset leader and $|C_{i}|=2m$ except $i\in X_{1}\cup X_{2}\cup X_{3}\cup X_{4}\cup X_{5}$, where \begin{center}
				$\begin{cases}
					X_{1}=\{aq^{t}+b:1\leq b\leq a\leq \frac{q-1}{2}\},\\
					X_{2}=\{aq^{t}-b:1\leq b< a\leq \frac{q-1}{2}\},\\
					X_{3}=\{aq^{t}+\frac{q^{t}-1}{2}+b,aq^{t}+\frac{q^{t}+1}{2}-b:1\leq b\leq a<\frac{q-1}{2}\},\\
					X_{4}=\{a(q^{t}+1)+\frac{q^{t}+1}{2}:0\leq a<\frac{q-1}{2}\},\\
					X_{5}=\{i:\frac{q^{t+1}-q^{2}}{2}+1\leq i\leq \frac{q^{t+1}-1}{2} \ and \ q\nmid i\}.
				\end{cases}$
			\end{center}
		\end{itemize}
		\begin{proof}
			We only prove the conclusion (2), as (1) can be similarly obtained by Lemma \ref{le4}. By Lemma \ref{le2}, we have $i\in \MinRep_{n}$ with $\frac{q^{t}+1}{2}\leq i\leq \frac{q^{t+1}-1}{2}$ and $i\not\equiv0\pmod{q}$ except that $2i\in A_{1}\cup A_{2}\cup A_{3}$, where $A_{1},A_{2},A_{3}$ are defined in the Lemma \ref{le13}.
			
			We divide $i$ into three cases to prove our results.
			\begin{itemize}
				\item[$\bf{(i)}$] If $2i\in A_{1}$, then $2i=a_{t}q^{t}+a_{0}$, where $1\leq a_{0}\leq a_{t}\leq q-1$.\\
				If $2\mid a_{t}$, then $2\mid a_{0}$. Denote that $a_{t}=2a$ and $a_{0}=2b$, then $i=aq^{t}+b$, where $1\leq b\leq a\leq \frac{q-1}{2}$. That is $X_{1}$.\\
				If $2\nmid a_{t}$, then we can assume that $a_{t}=2a+1$, where $0\leq a<\frac{q-1}{2}$. Note that $2i=a_{t}(q^{t}-1)+a_{t}+a_{0}$, then $a_{t}+a_{0}\equiv0\pmod {2}$, then we can assume that $a_{0}=2b-1$, where $b\geq 1$. Since $a_{t}-a_{0}=2(a-b)+2\geq0$, then $a-b\geq -1$. We will continue our discussion by distinguishing between the following two subcases.
				\begin{itemize}
					\item[-]If $a-b\geq 0$, then we have $1\leq b\leq a<\frac{q-1}{2}$ and $2i=(2a+1)q^{t}+2b-1$, i.e., $\{ aq^{t}+\frac{q^{t}-1}{2}+b:1\leq b\leq a<\frac{q-1}{2}\}\subset X_{3}$.
					\item[-]If $a-b=-1$, then we have $0\leq a<\frac{q-1}{2}$ and $2i=(2a+1)q^{t}+2a+1$, i.e., $\{ a(q^{t}+1)+\frac{q^{t}+1}{2}:0\leq a<\frac{q-1}{2}\}= X_{4}$.
				\end{itemize}
				\item[$\bf{(ii)}$] If $2i\in A_{2}$, then we can get $i\in X_{2}\cup \{ aq^{t}+\frac{q^{t}+1}{2}+b:1\leq b\leq a<\frac{q-1}{2}\}$ by the same way of $\bf{(i)}$.
				\item[$\bf{(iii)}$] If $2i\in A_{3}$, then we have $q^{t+1}-q^{2}+1\leq 2i\leq q^{t+1}$ and $q\nmid 2i$, i.e., $\{i:\frac{q^{t+1}-q^{2}}{2}+1\leq i\leq \frac{q^{t+1}-1}{2},\  q\nmid i\}=X_{5}$.
			\end{itemize}
			
			By Lemmas \ref{le2} and \ref{le13}, we have $|C_{i}|=2m$ for any $i\in \MinRep_{n}$ with $\frac{q^{t}+1}{2}\leq i\leq \frac{q^{t+1}-1}{2}$. Thus we complete the proof.
		\end{proof}
	\end{lemma}
	\begin{theorem}\label{th7}
		Let $n=\frac{q^{m}+1}{2}$ and $\epsilon=\lceil (\delta-1)(1-q^{-1})\rceil$. Then we have 
		\begin{itemize}
			\item[(1)]If $m=2t\geq 4$ and $\frac{q^{t}+1}{2}\leq \delta-1\leq q^{t}$, then LCD BCH code $\C_{(n,q,\delta,1)}$ has parameters $[\frac{q^{m}+1}{2},k,d\geq \delta]$ and $\C_{(n,q,\delta+1,0)}$ has parameters $[\frac{q^{m}+1}{2},k-1,d\geq 2\delta]$, where $k=n-2m(\epsilon-1)$.
			\item[(2)]If $m=2t\geq 8$ and $q^{t}+1\leq \delta-1\leq \frac{q^{t+1}-1}{2}$, where $q\geq 5$. Then LCD BCH code $\C_{(n,q,\delta,1)}$ has parameters $[\frac{q^{m}+1}{2},k,d\geq \delta]$ and $\C_{(n,q,\delta+1,0)}$ has parameters $[\frac{q^{m}+1}{2},k-1,d\geq 2\delta]$, where
			$$k=\begin{cases}
				n-2m(\epsilon-2a^{2}),& \text{if}\ a(q^{t}+1)< \delta\leq a(q^{t}-1)+\frac{q^{t}+1}{2};\\
				n-2m(\epsilon-2a^{2})+2m(\delta-[a(q^{t}-1)+\frac{q^{t}+1}{2}]),& \text{if}\ a(q^{t}-1)+\frac{q^{t}+1}{2}< \delta\leq a(q^{t}+1)+\frac{q^{t}+1}{2};\\
				n-2m(\epsilon-2a^{2}-2a-1),& \text{if}\ a(q^{t}+1)+\frac{q^{t}+1}{2} <\delta\leq (a+1)(q^{t}-1)+1;\\
				n-2m(\epsilon+(a+1)q^t-\delta-2a^{2}-3a-1),& \text{if}\ (a+1)(q^{t}-1)+1 <\delta\leq (a+1)(q^{t}+1),
			\end{cases}$$
			and $1\leq a<\frac{q-1}{2}$.
		\end{itemize}
		\begin{proof}
			From Lemma \ref{le14}, we can obtain the results following the same way of Theorem \ref{th3}.
		\end{proof}
	\end{theorem}
	
	\begin{example}
		Based on the above theorems, we have the following examples.
		\\$\bullet$ Let $(q,m)=(3,4)$, we have $n=\frac{q^{4}+1}{2}=41$. Then LCD BCH code $\C_{(n,q,9,0)}$ has parameters $[41,8,22]$, the code is distance-optimal according to the tables of best codes known in (\cite{RefJ10}).
		\\$\bullet$ Let $(q,m)=(5,4)$, we have $n=\frac{q^4+1}{2}=313$. Then LCD BCH code $\C_{(n,q,40,0)}$ has parameters $[313,96,\geq 78]$, the dimension is consistent with what was given in Theorem \ref{th7}.
	\end{example}
	\subsection{The the case of $\lambda=q+1$}
	In this subsection, we assume that $n=\frac{q^{m}+1}{q+1}$ and denote the $q$-cyclotomic coset modulo $n$ of representative $s$ as $C_{s,n}$, where $m=2t+1\geq 5$ is an integer.
	\begin{lemma}(\cite{RefJ22},\cite{RefJ24})\label{le15}
		Let $n=\frac{q^{m}+1}{\lambda}$ and $2\nmid m$, then 
		\begin{itemize}
			\item[(a)] If $\lambda=2$ and $\ q\equiv3\pmod{4}$, we have $\delta_{1,n}=\frac{q^{m}+1}{4}$ with $|C_{\delta_{1,n}}|=1$ and  $\delta_{2,n}=\frac{(q-1)q^{m-1}}{2}-\frac{q^{m}+1}{4}$ with $|C_{\delta_{2,n}}|=2m$.
			\item[(b)] If $\lambda=q+1$ and $1\leq a \leq q^{\frac{m-1}{2}}$ with $a\equiv0\pmod q$, then $a$ is a coset leader with $|C_{a}|=2m$ except that $a=\frac{q^{\frac{m+1}{2}}-(-1)^{\frac{m+1}{2}}}{q+1}$. 
		\end{itemize}	
	\end{lemma}
	\begin{lemma}\label{le16}
		If $m=2t+1\geq 5$, let $i$ be an integer with $q^{t}+1\leq i\leq  \frac{2q^{t+1}+2q-1}{q+1}$ and $i\not\equiv0\pmod{q}$.
		Then $i$ is a coset leader and $|C_{i}|=2m$ except $i\in Y_{1}\cup Y_{2}$. 
		\begin{itemize}
			\item[(1)] If $2\nmid t$, then
			\begin{center}
				$\begin{cases}
					Y_{1}=\{2\frac{q^{t+1}-1}{q+1}+c:c=-1,0\}, \  &\text{if}\quad q=3;\\
					Y_{1}=\{2\frac{q^{t+1}-1}{q+1}+c:c=\pm1,0\}, \ &\text{if}\quad q\geq 5.
				\end{cases}$ $\quad$ and $\quad Y_{2}=\{\frac{q^{t+1}+2q^{t}+1}{q+1}\}$.
			\end{center} 
			\item[(2)] If $2\mid t$, then
			\begin{center}
				$\begin{cases}
					Y_{1}=\{2\frac{q^{t+1}+1}{q+1}+c:c=-1,0\}, \  &\text{if}\quad q=3;\\
					Y_{1}=\{2\frac{q^{t+1}+1}{q+1}+c:c=\pm1,0\}, \  &\text{if}\quad q\geq 5,
				\end{cases}$ $\quad$ and $\quad Y_{2}=\{\frac{q^{t+1}+2q^{t}-1}{q+1}\}$.
			\end{center}
		\end{itemize}
		\begin{proof}
			From Lemmas \ref{le2} and \ref{le4}, we can obtain the results following the same way of Lemma \ref{le12}.
		\end{proof}
	\end{lemma}
	\begin{theorem}\label{th8}
		If $m=2t+1\geq 5$ and $q\geq 5$, let $\epsilon=\lceil (\delta-1)(1-q^{-1})\rceil$ and $$\begin{cases}
			A=\frac{q^{t+1}+2q^{t}+1}{q+1},\ B=2\frac{q^{t+1}-1}{q+1}-1, \quad  &\text{if}\quad 2\nmid t;\\
			A=\frac{q^{t+1}+2q^{t}-1}{q+1},\ B=2\frac{q^{t+1}+1}{q+1}-1, \quad  &\text{if}\quad 2\mid t.
		\end{cases}$$
		Then LCD BCH code $\C_{(n,q,\delta,1)}$ has parameters $[\frac{q^{m}+1}{q+1},k,d\geq \delta]$ and $\C_{(n,q,\delta+1,0)}$ has parameters $[\frac{q^{m}+1}{q+1},k-1,d\geq 2\delta]$, where $$k=
		\begin{cases}
			n-2m(\epsilon-1), \quad  &\text{if}\quad q^{t}+2\leq \delta\leq A;\\	
			n-2m(\epsilon-2), \quad  &\text{if}\quad A+1\leq \delta\leq B.
		\end{cases}$$
		\begin{proof}
			From Lemmas \ref{le15} and \ref{le16}, we can obtain the results by the same way of Theorem \ref{th3}.
		\end{proof}
	\end{theorem}
	
	Next we will investigate the largest and second largest coset leaders modulo $n$ when $q=3$, and obtain parameters of the BCH code $\C_{(q,n,\delta_{i,n},b)}$.
	\begin{lemma}\label{le17}
		Let $n=\frac{3^{m}+1}{4}$ and $1\leq a<n$ be an integer, then $a$ is a coset leader of $C_{a}$ modulo $\frac{3^{m}+1}{4}$ if and only if $2a$ is a coset leader of $C_{2a}$ modulo $\frac{3^{m}+1}{2}$. Moreover, $|C_{a}|=| C_{2a}|$.
		\begin{proof}
			From Lemma \ref{le2}, we can get the result directly.
		\end{proof}	
	\end{lemma}
	\begin{lemma}\label{le18}
		Let $n=\frac{3^{m}+1}{4}$, where $m\geq 5$ is an odd integer. Then 
		$$\delta_{1}=\frac{3^{m-1}-1}{8}, \quad \delta_{2}=\frac{3^{m-1}-1}{8}-2.$$
		are the first and second largest coset leaders modulo $n$. Moreover, $| C_{\delta_{1}}| =| C_{\delta_{2}}|=2m$.
		\begin{proof}
			In order to prove this, we divide it into two steps.
			
			$\bf{Step 1}$. We claim that $\delta_{1},\delta_2\in \MinRep_{n}$. From Lemma \ref{le15}, we have $\delta_{2,2n}=\frac{3^{m-1}-1}{4}$. Note that $2\mid \frac{3^{m-1}-1}{4}$, then $\delta_{1}=\frac{3^{m-1}-1}{8}\in \MinRep_{n}$ and $|C_{\delta_{1}}| =2m$ by Lemma \ref{le17}.	
			
			Note that $\delta_{2}\in \MinRep_{n}$ if and only if  $\delta=4\delta_{2}=\frac{3^{m-1}-1}{2}-8\in \MinRep_{3^{m}+1}$ by Lemma \ref{le2}. It is clear that $C_{\delta}=\{\big[q^{k}\delta\big]_{3^{m}+1},3^{m}+1-\big[q^{k}\delta\big]_{3^{m}+1}: 0\leq k\leq m-1\}$, then we need to prove that $\big[q^{k}\delta\big]_{3^{m}+1}-\delta\geq 0$ and $3^{m}+1-\big[q^{k}\delta\big]_{3^{m}+1}-\delta\geq 0$ for any $0\leq k\leq m-1$. It is easy to get that
			$$\delta=\frac{3^{m-1}-1}{3-1}-8=3^{m-2}+3^{m-3}+\cdots+3^{3}+3+2,$$
			$$3^{m}+1-\big[q^{0}\delta\big]_{3^{m}+1}-\delta=2\cdot3^{m-1}+18>0.$$
			For $k=1$, we have
			\begin{center}	 		
				$\begin{aligned}
					\big[q\delta\big]_{3^{m}+1}-\delta&=3^{m-1}+3^{m-2}+\cdots+3^{1}+(3-3^{3})-\delta\\&=3^{m-1}-2\cdot3^{2}+1>0,\\
					3^{m}+1-\big[q\delta\big]_{3^{m}+1}-\delta&=3^{m}+1-(2\delta+3^{m-1}-2\cdot3^{2}+1)\\&=3^{m-1}+3^{3}+3^{2}-1>0.
				\end{aligned} $
			\end{center}
			For $2\leq k\leq m-3$, we have  		
			$$\begin{aligned}
				\big[q^{k}\delta\big]_{3^{m}+1}-\delta&=\sum^{m-1}_{t=k}3^{t}-3^{k+2}+3^{k}-\sum_{t=0}^{k-2}3^{t}-\delta\\
				&\geq3^{m-2}+2\cdot3^{m-3}-\sum_{t=0}^{m-5}3^{t}-\delta\\
				&=3^{m-4}+3^{2}>0,
			\end{aligned} $$
			$$\begin{aligned}
				3^{m}+1-\big[q^{k}\delta\big]_{3^{m}+1}-\delta&=3^{m}+1-(\sum^{m-1}_{t=k}3^{t}-3^{k+2}+3^{k}-\sum_{t=0}^{k-2}3^{t}+\delta)\\
				&\geq3^{m}+1-(\sum^{m-1}_{t=2}3^{t}-3^{4}+3^{2}-1+\delta)\\
				&=3^{m-1}+3^{4}-2>0.
			\end{aligned} $$
			For $ k= m-2$, we have 
			\begin{center}
				$\begin{aligned}
					\big[q^{m-2}\delta\big]_{3^{m}+1}-\delta&=3^{m-1}+2\cdot3^{m-2}-\sum^{m-4}_{t=1}3^{t}-\delta\\&=3^{m-1}+3^{m-3}+3^{2}>0,\\
					3^{m}+1-\big[q^{m-2}\delta\big]_{3^{m}+1}-\delta&=3^{m}+1-(2\delta+3^{m-1}+3^{m-3}+3^{2})\\&=3^{m-1}-3^{m-3}+3^{2}-1>0.
				\end{aligned} $
			\end{center}
			For $ k= m-1$, we have 
			$$\begin{aligned}
				\big[q^{m-1}\delta\big]_{3^{m}+1}-\delta&=2\cdot3^{m-1}-\sum^{m-3}_{t=2}3^{t}-1-\delta\\&=3^{m-1}+3^{m-2}+3^{2}+2\cdot3+1>0,\end{aligned} $$
			$$\begin{aligned}
				3^{m}+1-\big[q^{m-1}\delta\big]_{3^{m}+1}-\delta&=3^{m}+1-(2\delta+3^{m-1}+3^{m-2}+3^{2}+2\cdot3+1)\\&=2\cdot3^{m-3}+5>0.
			\end{aligned} $$
			Therefore, $\delta\in \MinRep_{3^{m}+1}$ and $| C_{\delta}|=2m$, i.e., $\delta_{2}\in \MinRep_{n}$ and $| C_{\delta_{2}}|=2m$ by Lemma \ref{le2}.
			
			$\bf{Step 2}$. We claim that $\delta_{1,n}=\delta_{1}$ and $\delta_{2,n}=\delta_{2}$. Suppose there exists an integer $a$ such that $a\in \MinRep_{n}$ and $\delta_{1}<a< n$, which implies that $2a\in \MinRep_{\frac{q^m+1}{2}}$ with $\delta_{2,\frac{q^m+1}{2}}=\frac{3^{m-1}-1}{4}<2a<\frac{q^{m}+1}{2}$ by Lemma \ref{le17}, then $2a=\delta_{1,\frac{q^m+1}{2}}=\frac{3^{m}+1}{4}$ by Lemma \ref{le15}, which contradicts to the fact that $2\nmid\frac{3^{m}+1}{4}$. Therefore, $\delta_{1,n}=\delta_1$.
			
			Note that $3\mid \frac{3^{m-1}-1}{8}-1$ and $\frac{3^{m-1}-1}{8}-2\in\MinRep_{n}$, then we have $\delta_{2,n}=\frac{3^{m-1}-1}{8}-2$.
			Thus we complete the proof.
		\end{proof}
	\end{lemma}
	According to the previous results on the coset leaders, we give the following theorem.
	\begin{theorem}\label{th9}
		Let $n=\frac{3^{m}+1}{4}$, where $m=2t+1\geq 5$, then LCD BCH code  $\C_{(\frac{3^{m}+1}{4},3,\delta,1)}$ has parameters $[n,2m+1,d\geq \delta_{2,n}]$ and $\C_{(\frac{3^{m}+1}{4},3,\delta+1,0)}$ has parameters $[n,2m,d\geq 2\delta_{2,n}] $, where $\delta_{2,n}+1\leq \delta\leq \delta_{1,n}$.
		\begin{proof}
			From Lemma \ref{le18}, we can obtain the results directly.
		\end{proof}
	\end{theorem}
	\begin{example}
		Based on the above theorems, we have the following examples.
		\\$\bullet$ Let $(q,m)=(3,5)$, we have $n=\frac{q^{5}+1}{q+1}=61$. The LCD BCH code $\C_{(n,q,3,0)}$ has parameters $[61,50,6]$, the code is distance-optimal according to the tables of best codes known in (\cite{RefJ10}).
		\\$\bullet$ Let $(q,m)=(5,5)$, we have $n=\frac{q^5+1}{q+1}=521$. The LCD BCH code $\C_{(n,q,29,0)}$ has parameters $[521,320,\geq 56]$, the dimension is consistent with what was given in Theorem \ref{th8}.
	\end{example}
	\section{Conclusions}\label{set6} The main contributions of this paper are as follows:
	\begin{itemize}
		\item For the LCD BCH codes of length $n=q^{m}+1$ with $m=4t+2\geq 2$, we found the first few largest coset leaders $\delta_{i,n}$, and investigated the parameters of $\C_{(n,q,\delta,1)}$ and $\C_{(n,q,\delta+1,0)}$ with $ \delta_{i,n}+1\leq\delta\leq n$ (see Theorems \ref{th1} and \ref{th2}).
		
		\item For the LCD BCH codes of length $n=\frac{q^{m}+1}{\lambda}$ with $q+1>\lambda\geq2$. When $m=2t+1\geq 5$ and $1\leq \delta-1\leq  \frac{2q^{t+1}+2q-1}{\lambda}$, we completely provided the dimensions of $\C_{(n,q,\delta,1)}$ and $\C_{(n,q,\delta+1,0)}$ (see Theorems \ref{th3} and \ref{th4}). When $\lambda=2$ and $m=2t\geq 8$, we provided the dimensions of $\C_{(n,q,\delta,1)}$ and $\C_{(n,q,\delta+1,0)}$ for $\frac{q^{t}+1}{2}\leq \delta-1\leq \frac{q^{t+1}-1}{2}$ (see Theorem \ref{th7}).
		
		\item For the LCD BCH codes of length $n=\frac{q^{m}+1}{q+1}$ with $m=2t+1\geq5$, we presented the dimensions of $\C_{(n,q,\delta,1)}$ and $\C_{(n,q,\delta+1,0)}$ for $q^{t}+1\leq \delta-1\leq \frac{2q^{t+1}+2q-1}{q+1}$ (see Theorem \ref{th8}). For $q=3$, the first two largest coset leaders $\delta_{1,n}$ and $\delta_{2,n}$ modulo $n$ were presented, and the parameters of $\C_{(n,q,\delta,1)}$ and $\C_{(n,q,\delta+1,0)}$ with $\delta_{2,n}+1\leq\delta\leq \delta_{1,n}$ were studied (see Theorem \ref{th9}).
	\end{itemize}
	\section*{Acknowledgements}
	The work was supported by the National Natural Science Foundation of China (Nos.12271137,12271335,62201009) and Natural Science Foundation of Anhui Province (No.2108085QA06).
	
\end{document}